\documentclass{article}
\pdfoutput=1

\usepackage[utf8]{inputenc} 
\usepackage[T1]{fontenc}    
\PassOptionsToPackage{hyphens}{url} 
\usepackage[hyperfootnotes=false]{hyperref}
\usepackage{booktabs}       
\usepackage{amsfonts}       
\usepackage{nicefrac}       
\usepackage{microtype}      
\usepackage[table]{xcolor}  
\usepackage{breqn}
\usepackage[hang,flushmargin]{footmisc}
\usepackage{algorithm}
\usepackage{algpseudocode}
\usepackage{graphicx}
\usepackage{enumerate}
\usepackage{caption}
\usepackage{tcolorbox}      
\usepackage{array}           
\usepackage{subcaption}
\usepackage{comment}
\usepackage[export]{adjustbox}
\usepackage{mathtools}
\usepackage{amsthm}
\usepackage{authblk}
\usepackage{fullpage}

\usepackage{enumitem}

\usepackage[round]{natbib}

\usepackage{tikz}
\usetikzlibrary{positioning}
\usetikzlibrary{calc}
\usetikzlibrary{fit}
\usepackage[HTML]{xcolor}
\usepackage{tcolorbox}

\usepackage{Definitions}
\usepackage{dsfont}

\definecolor{mycommentcolor}{HTML}{4f9739}
\algrenewcommand\algorithmiccomment[1]{\hfill\textcolor{mycommentcolor}{\(\triangleright\) #1}}

\newlength{\inlineheight}
\newcommand{\UnnumberedFootnote}[1]{{\def\thefootnote{}\footnote{#1}
\addtocounter{footnote}{-1}}}

\algnewcommand{\LineComment}[1]{\State \textcolor{mycommentcolor}{\(\triangleright\) #1}}

\title{Auditing Pay-Per-Token in Large Language Models}

\author{Ander Artola Velasco$^{\S}$}
\author{Stratis Tsirtsis$^{*,\dagger}$}
\author{Manuel~Gomez-Rodriguez$^{\S}$}
\affil{$^{\S}$Max Planck Institute for Software Systems, Kaiserslautern, Germany \\
\{avelasco, manuel\}@mpi-sws.org}

\affil{$^{\dagger}$Hasso Plattner Institute, Potsdam, Germany \\ stratis.tsirtsis@hpi.de}

\date{}

\begin{document}

\maketitle

\UnnumberedFootnote{$^{*}$The author contributed to this work during his doctoral studies at the Max Planck Institute for Software Systems.}

\begin{abstract}
Millions of users rely on a market of cloud-based services to obtain access to state-of-the-art large language models.
%
However, it has been very recently shown that the de facto \emph{pay-per-token} pricing mechanism used by providers creates a financial incentive for them to strategize and misreport the (number of) tokens a model used to generate an output.
%
%
In this paper, we develop an auditing framework based on martingale theory that enables a trusted third-party auditor who sequentially queries a provider to detect token misreporting.
%
%
Crucially,
we show that our framework is guaranteed to \emph{always} detect token misreporting, regardless of the provider's (mis-)reporting policy, 
and not falsely flag  
a faithful provider as unfaithful with high probability.
%
To validate our auditing framework, we conduct experiments across a wide range of (mis-)reporting policies using several large language models from the \texttt{Llama}, \texttt{Gemma} and \texttt{Ministral} families, and input prompts from a popular crowdsourced benchmarking platform.
The results show that our framework detects an unfaithful provider after observing fewer than $\sim$$70$ reported outputs, 
while maintaining the probability of falsely flagging a faithful provider below $\alpha = 0.05$.
\end{abstract}

\section{Introduction}
\label{sec:intro}
%
%
State-of-the-art large language models (LLMs) require a vast amount of resources, often involving hundreds or thousands of GPUs or TPUs, along with specialized infrastructure to handle massive parallel computations~\citep{narayanan2021efficient, samsi2023words, jiang2024megascale}. 
As a consequence, most (enterprise) users cannot operate them locally, and instead rely on a rapidly growing market of cloud-based providers that offer LLMs-as-a-service~\citep{chen2023acceleratinglargelanguagemodel, snell2024scalingllmtesttimecompute,Pais2022, patel2024cloudplatformsdevelopinggenerative}.

%
%
In a typical LLM-as-a-service, a user submits a prompt to the provider via an application programming interface (API).
Then, the provider feeds the prompt into an LLM running on their own hardware, which generates a sequence of tokens\footnote{Tokens are units that make up sentences and paragraphs, such as (sub-)words, symbols and numbers.} as an output using a (non-deterministic) generative process.
Finally, the provider shares the output with the user and charges them based on a simple pricing mechanism---a fixed price per token.\footnote{\url{https://ai.google.dev/gemini-api/docs/pricing}, \url{https://openai.com/api/pricing/}.}

%
%
Very recently,~\cite{velasco2025llmoverchargingyoutokenization} have argued, both theoretically and empirically, that the above pricing mechanism creates a financial incentive for providers to strategize. 
Their key observation is that, in the interaction between a user and a provider, there is an asymmetry of information~\citep{milgrom1987informational,rasmusen1989games,mishra1998information}, which enables a situation known in economics as moral hazard~\citep{holmstrom1979moral}. 
In particular, the provider observes the entire generative process used by the LLM to generate an output, whereas the user only observes and pays for the output shared with them by the provider. 
As a consequence, the provider has the opportunity to misreport the number of tokens in an output to increase their profit, at the expense of the user, and the user cannot know whether a provider is overcharging them.

%
%
The core of the problem lies in the fact that the tokenization of a string is not unique, and an LLM can in principle generate different tokenizations of the same string~\citep{geh-etal-2024-signal, cao-rimell-2021-evaluate, chirkova-etal-2023-marginalize}.
For example, consider that the user submits the prompt 
``\texttt{Where does the next AISTATS take place?}'' to the provider, the provider feeds it into an LLM, and the model generates the output ``$|\texttt{Tang}|\texttt{ier}|\texttt{,}|\texttt{ Morocco}|$'' consisting of four tokens. 
Exploiting the asymmetry of information, a self-serving provider could simply claim that the LLM generated the tokenization ``$|\texttt{Tang}|\texttt{ier}|\texttt{,}|\texttt{ }|\texttt{Mor}|\texttt{oc}|\texttt{co}|$'' and overcharge the user for seven tokens instead of four! 

%
%
In this paper, we consider a forward-looking yet realistic\footnote{\label{fn:AI-ACT}According to Article 74(13) of the EU AI Act, ``market surveillance authorities shall be granted access to the source code of the high-risk AI system [...] when testing or auditing procedures and verifications based on the data and documentation provided by the provider have been exhausted or proved insufficient.''} scenario in which, to eliminate the incentive for providers to engage in misreporting, they are required to share information about the generative process with a third-party trusted auditor who can verify they are faithful.

\xhdr{Our contributions}
%
%
We start by formalizing the problem of auditing for token misreporting by a provider as a sequential hypothesis test. In doing so, we consider that the third-party trusted auditor has access to the next-token probability distribution of the model served by the provider.
Then, building upon this formalization, we introduce an auditing framework that uses a sequential statistical test based on the theory of martingales~\citep{ramdas2023gametheoreticstatisticssafeanytimevalid} to detect token misreporting.
%
Along the way, we develop a novel, unbiased, and efficient estimator of the average length of the token sequences used by a model to encode any given output string, which our statistical test uses and may be of independent interest.
Further, we provide sufficient conditions under which
%
our framework is guaranteed to eventually detect an unfaithful provider, regardless of 
their (mis-)reporting policy, 
and not falsely flag faithful providers as unfaithful with high probability.

%
%
To validate our auditing framework, we conduct experiments using several large language models from the \texttt{Llama}, \texttt{Gemma} and \texttt{Ministral} families, and input prompts from a popular crowdsourced benchmarking platform.
The results show that,
%
for several 
\mbox{(mis-)reporting} policies introduced in prior work~\citep{velasco2025llmoverchargingyoutokenization},
our auditing framework detects an unfaithful provider after observing just up to $\sim$$70$ reported outputs,
while maintaining the probability of falsely flagging a faithful provider below a prespecified threshold $\alpha=0.05$.\footnote{The code for our experiments is publicly available at \url{https://github.com/Human-Centric-Machine-Learning/token-audit}.}

%
%
\xhdr{Further related work}
Our work builds upon further related work on the economics of LLMs-as-a-service, tokenization, and sequential statistical tests.

%
%
Within the rapidly growing literature on the economics of LLMs-as-a-service~\citep{la2024language,mahmood2024pricing,laufer2024fine,cai2025are,saig2024incentivizing,bergemann2025economicslargelanguagemodels,velasco2025llmoverchargingyoutokenization}, there has been increasing interest in the ways in which providers may strategically act at the expense of users.
%
%
%
Within this literature, our work is most closely related to a line of work on algorithmic auditing~\citep{bourree2025robust,sun2025coincountinginvisiblereasoning,wang2025predictiveauditinghiddentokens}, which has focused on detecting whether an LLM provider is unfaithful about the model they serve or the token counts during hidden reasoning steps.
%
%
%
%
Yet, this line of work has largely overlooked the possibility that an LLM provider may be unfaithful about the tokenization of the outputs---a threat that has not been studied until very recently~\citep{velasco2025llmoverchargingyoutokenization}.

Multiple lines of empirical evidence have shown that tokenization plays a central role in developing and analyzing LLMs~\citep{geh-etal-2024-signal,giulianelli-etal-2024-proper,geh2025adversarial,petrov2023language, ovalle2024tokenizationmattersnavigatingdatascarce,chatzi2024counterfactual}.
Consequently, there have been numerous efforts to better understand and improve byte-pair encoding (BPE), the tokenization algorithm most commonly used in LLMs~\citep{bostrom-durrett-2020-byte, zouhar2023formal,lian2024scaffold, sennrich2015neural,lian2024lbpe}. 
However, this line of work has not studied the economic implications of tokenization (in the context of LLMs-as-a-service), which is the main focus of our work.

%
Our work also builds upon the active and expanding body of research on sequential statistical testing with martingales and e-values~\citep{ramdas2023gametheoreticstatisticssafeanytimevalid, ramdas2025hypothesistestingevalues, waudbysmith2025universallogoptimalitygeneralclasses}, which has derived stronger guarantees than classical testing approaches and has been successfully applied to a wide range of statistical problems~\citep{shin_ramdas_rinaldo_2024, 10.1109/TIT.2023.3305867, pmlr-v238-xu24a}. In a concurrent work,~\citet{gauthier2026bettingequilibriummonitoringstrategic} obtain results similar to ours at a technical level. However, they focus on detecting deviations from Nash equilibria in multi-agent games. To the best of our knowledge, we are the first to use techniques based on e-values in the context of auditing LLMs-as-a-service.

\section{Auditing for Token Misreporting as a Sequential Hypothesis Test}
\label{sec:model}
%
We model the process of auditing a provider for token misreporting as a sequential interaction between the provider serving an LLM $\Mcal$ and an auditor. At each time step $i=1,2,\dots$, the auditor selects a prompt $Q \sim P^Q$ from a fixed prompt distribution and queries the provider for a response to $Q$.\footnote{We denote random variables with capital letters ($X$) and their realizations with lower case letters ($x$).} 
The provider then generates a sequence of tokens $\Tb\in \Vcal^*$ by autoregressively sampling one token at a time, 
where $\Vcal^*$ is the set of finite sequences of tokens in the vocabulary of tokens $\Vcal$ used by the LLM $\Mcal$.

%
Importantly, since only the provider observes the sequence $\Tb$, they have the capacity to report a different sequence of tokens $\widetilde{\Tb} \sim\pi(Q,\Tb)$ 
%
using a (non-deterministic) reporting policy $\pi$, which is unknown to the auditor.\footnote{In principle, a provider could use, at each time step $i$, a different reporting policy. We discuss this possibility in Section~\ref{sec:discussion}.}
%
%
%
While, in principle, a provider can choose any reporting policy they wish, we narrow our focus to reporting policies that misreport the tokens in $\Tb$ \emph{while preserving its string-level representation}, similarly as in~\citet{velasco2025llmoverchargingyoutokenization}; that is, $\texttt{str}(\widetilde{\Tb}) = \texttt{str}(\Tb)$ for any $\widetilde{\Tb} \sim \pi(Q, \Tb)$, where $\texttt{str} \colon \Vcal^* \rightarrow \Scal$ maps a sequence of tokens to the respective string, and $\Scal$ denotes the set of all possible strings.
Here, it is also important to note that, for a given prompt $q$ and sequence of tokens $\tb$, the provider can only obtain a financial benefit if
\begin{equation}
    \EE_{\widetilde{\Tb}\sim \pi(q, \tb)} \left[ \len(\widetilde{\Tb}) 
    \right] \geq 
    \len(\tb),
\end{equation}
where $\texttt{len}\,\colon \Vcal^* \to \mathbb{N}$ denotes the number of tokens in a given sequence. This is because, under the de facto standard pay-per-token pricing, the price charged to users for a reported output sequence $\widetilde{\Tb}$ increases linearly with its length. 
Therefore, we can naturally characterize the financial benefit the provider obtains from misreporting by measuring the average number of additional tokens in $\widetilde{\Tb}$ compared to $\Tb$ across all prompts $q$ sampled from $P^{Q}$, which we refer to as the \emph{misreporting intensity} $\Ical(\pi)$, \ie,
\begin{equation}\label{eq:misreporting-intensity}
        \Ical(\pi) = \EE_{Q \sim P^{Q}} \left [ \EE_{\Tb \sim P^{\Mcal}(\cdot \,|\, q)} \left[ \EE_{\widetilde{\Tb}\ \sim\ \pi(q, \tb)} \left[\Delta_{\pi}(\tb))\right]  \,|\, \Tb=\tb \right] \,|\,  Q=q \right],
\end{equation}
where
\begin{equation}\label{eq:delta}
    \Delta_{\pi}(\tb) = \EE_{\widetilde{\Tb}\ \sim\ \pi(q, \tb)} \Big[
        \\\mathrm{\texttt{len}}(\widetilde{\Tb})\Big] - \texttt{len}(\tb),
\end{equation}
%
and $P^{\Mcal}(\cdot \,|\, q)$ denotes the probability distribution of the sequences $\Tb$ generated by the LLM $\Mcal$ in response to the prompt $q$.
Intuitively, the misreporting intensity measures the severity of the misreporting 
and hence, for an auditor, it is more critical to detect misreporting policies with high intensity---a policy satisfying $\Ical(\pi) \approx 0$ would minimally harm the user. Moreover, as we will demonstrate later, both theoretically and empirically, the difficulty of detecting if a provider using an unknown policy $\pi$ is engaging in misreporting is fundamentally determined by $\Ical(\pi)$.


%
Under the above characterization, auditing for token misreporting can be framed as a (sequential) hypothesis test on the misreporting intensity, where an auditor (sequentially) gathers sufficient statistical evidence to conclude that the lengths of the reported sequences $\widetilde{\Tb}$ do not \emph{match}, in expectation, the lengths of the sequences $\Tb$ generated by the model.
More concretely, we can define the following null and alternative hypotheses:
\begin{equation}\label{eq:hypo-test}
    \begin{dcases}
        H_0=\left\{ \pi_0  \right\} &\quad \quad (\mathrm{null})\\
        H_1=\left\{ \pi : \Ical(\pi)>0  \right\} &\quad \quad (\mathrm{alternative})
\end{dcases},
\end{equation}
where $\pi_0$ is the policy that faithfully reports tokens in $\Tb$, \ie, $\pi_0(Q,\Tb) \coloneqq \Tb$, with $\Ical(\pi_0)=0$, and note that the hypothesis $H_1$ includes any family of (non-deterministic) reporting policies, potentially of arbitrary sophistication. 
%
%
In the next section, we will develop a framework to test the above hypothesis in a setting in which the auditor has access to the next-token probabilities of the model.\footnote{Although this is necessary for our theoretical analysis, in practice, our auditing framework performs similarly in settings where the auditor has access to approximate values of next-token probabilities (see Appendix~\ref{app:robust}).} 
Such a setting fits a variety of real-world scenarios, for example, a scenario in which the provider serves an open-weight model, or a scenario in which the provider is required, by regulation, to grant trusted auditors access to a proprietary model (refer to Section~\ref{sec:discussion} for further discussion on this assumption).


\section{A Sequential Hypothesis Test Auditing Framework} 
\label{sec:test}
%
%
%
Our starting point is the key observation that, as long as the reporting policy $\pi$ preserves the string-level representation $\Sbb$ of the generated sequences $\Tb$, the inner expectation in the misreporting intensity defined by Eq.~\ref{eq:misreporting-intensity} can be expressed as follows:
\begin{equation}
\EE_{\Sbb \sim P^{\Mcal}(\cdot \given q)}  \Big[ \EE_{\Tb \sim P^{\Mcal}_{\sbb}(\cdot \given q)} \Big[ \Delta_{\pi}(\tb)  \given \Tb=\tb \Big] \given \Sbb=\sbb \Big], 
\end{equation}
where $\Delta_\pi(\tb)$ is as in Eq.~\ref{eq:delta}
%
and $P^{\Mcal}_{\sbb}(\cdot \given q)$ denotes the conditional distribution of the sequences $\Tb$ generated by the LLM $\Mcal$ in response to the prompt $q$ whose string-level representation matches $\sbb$.

As a consequence, auditing for token misreporting reduces to gathering sufficient (statistical) evidence to conclude that the lengths of the reported sequences $\widetilde{\Tb}$ do not \emph{match}, in expectation, the average length of the sequences $\Tb$ used by the LLM $\Mcal$ to encode the respective string $\sbb = \texttt{str}(\widetilde{\Tb})$, \ie, they do not match $\EE_{\Tb \sim P^{\Mcal}_{\sbb}(\cdot \given q)}\left[ \textnormal{\texttt{len}}(\Tb)\right]$.
In what follows, we will first introduce an efficient and unbiased estimator of the conditional average mentioned above for any given string $\sbb$, and then leverage this estimator to design a statistical (sequential) test to determine whether a provider is engaging in token misreporting, that is, to test $H_0$ against $H_1$.
%

%
%
%

\subsection{Estimating the Average Length of Token Sequences Encoding a Given String}

To estimate the average length $\EE_{\Tb \sim P^{\Mcal}_{\sbb}(\cdot \given q)}\left[ \textnormal{\texttt{len}}(\Tb)\right]$, 
we first modify the autoregressive generation process used by the LLM $\Mcal$ so that it always generates sequences of tokens $\Tb$ whose string-level representation satisfies $\texttt{str}(\Tb) = \sbb$, and then use this modified process to compute an unbiased Monte Carlo estimate of the average token sequence length.

Let $P^{\Mcal}(t \given q, \tb)$ denote the next-token probability that the LLM $\Mcal$ assigns to a token $t$ given a prompt $q$ and a partial output token sequence $\tb$. 
%
The modified generation process instead utilizes the following \emph{masked} next-token probability:
\begin{equation} \label{eq:locally-constrained}
    \widehat{P}^{\Mcal}_{\sbb}(t \given q, \tb):=\begin{cases}
        P^{\Mcal}(t \given q, \tb)/Z  & \text{if ${\tb} \shortmid t \models \sbb$} \\
        0 & \text{otherwise,}
    \end{cases}
\end{equation}
where ${\tb} \shortmid t \models \sbb$ indicates that the string-level representation of the concatenated sequence ${\tb} \shortmid t$ is a prefix of $\sbb$, and $Z= \sum_{t \in \Vcal:\ {\tb} \shortmid t \models \sbb} P^{\Mcal}(t \given q, \tb)$ is a normalization constant that ensures that the values $\widehat{P}^{\Mcal}_s(t \given q, \tb)$ lead to a valid probability distribution over the vocabulary $\Vcal$.
%
%
In words, the modified generation process simply masks any token that would lead to a string different than $\sbb$. 
%
%
Even though one could think otherwise, the distribution of token sequences $\widehat{P}_{\sbb}^\Mcal(\cdot \given q)$ generated using the above modified process does not in general match the conditional distribution $P^{\Mcal}_{\sbb}(\cdot \given q)$, as recently noted by~\cite{lipkin2025fastcontrolledgenerationlanguage}.
However, perhaps surprisingly, it can be used to efficiently construct an unbiased estimate of the average length $\EE_{\Tb \sim P^{\Mcal}_{\sbb}(\cdot \given q)}\left[ \textnormal{\texttt{len}}(\Tb)\right]$, as we show next.
%

Let $K \sim P^K$, where $P^K$ is a distribution with support over $\mathbb{N}$, and $\widehat{\Tb}_1, \dots, \widehat{\Tb}_K$ be $K$ independent samples from $\widehat{P}_{\sbb}^\Mcal(\cdot \given q)$, generated using the modified generation process mentioned above.
Moreover, define $R_0=0$ and, for $k\in\{1,\dots,K \}$, let $R_k$ be the weighted average
%
\begin{equation} \label{eq:rk}
    R_k=\frac{\sum_{j=1}^k w_j \cdot \textnormal{\texttt{len}}(\widehat{\Tb}_j)}{\sum_{j=1}^k w_j}, 
\end{equation} 
where $w_j = P^\Mcal(\widehat{\Tb}_{j} \given q) / \widehat{P}^{\Mcal}_{\sbb}(\widehat{\Tb}_{j}\given q)$ is the relative likelihood of the $j$-th sample between the original and the modified generation process of the model $\Mcal$.
Then, as formalized by Algorithm~\ref{alg:estimator} and Proposition~\ref{prop:unbiased-estimator}, we can efficiently construct an unbiased estimate of the average length $\EE_{\Tb \sim P^{\Mcal}_{\sbb}(\cdot \given q)}\left[ \textnormal{\texttt{len}}(\Tb)\right]$ using a weighted sum of the increments $R_{k} - R_{k-1}$:\footnote{All proofs can be found in Appendix~\ref{app:proofs}.}

%
%

\begin{proposition}\label{prop:unbiased-estimator}
Let $K \sim P^{K}$ and $R_{k}$ be defined by Eq.~\ref{eq:rk} for each $k \in \{0, \dots, K\}$. Then, it holds that:
    \begin{equation}
        \EE_{K\sim P^K,\, \widehat{\Tb}_{k} \sim \widehat{P}^{\Mcal}_{\sbb}(\cdot \given q)} 
        \left[     \sum_{k=1}^K\frac{R_k-R_{k-1}}{P^K(K\geq k)} \right]
        = \EE_{\Tb \sim P^{\Mcal}_{\sbb}(\cdot \given q)}\left[ \textnormal{\texttt{len}}(\Tb)\right].
    \end{equation}
\end{proposition}
In the next section, we will show how an auditor who sequentially queries the provider can leverage the above estimator to conclude that the provider is (not) engaging in misreporting.

\setlength{\textfloatsep}{10pt}
\begin{algorithm}[t]
\caption{It returns an unbiased estimate of the average length $\EE_{\Tb \sim P^{\Mcal}_{\sbb}(\cdot \given q)}\left[ \textnormal{\texttt{len}}(\Tb)\right]$.
}
\label{alg:estimator}
\begin{algorithmic}[1]

\algrenewcommand\algorithmicprocedure{\textbf{function}}

\Procedure{EstimateLength}{Prompt $q$, output string $\sbb$, next-token probability $P^{\Mcal}$, distribution $P^{K}$}

\State Sample $K \sim P^K$

\For{$k=1,\dots,K$}                         
    \State Sample $\widehat{\Tb}_k \sim \widehat{P}^{\Mcal}_{\sbb}(\cdot \given q, \tb)$ 
    \State $w_k \gets \frac{P^\Mcal(\widehat{\Tb}_{k}|q)}{\widehat{P}^{\Mcal}_{\sbb}(\widehat{\Tb}_{k}|q)}$ 
    \State $Z_k \gets \sum_{j=1}^k w_j$ 

\EndFor 
\State $R_0 \gets 0$
\For{$k=1,\dots,K$} 
    \State $w'_k \gets \frac{w_k}{Z_k}$ 
    \State $R_k \gets \sum_{j=1}^k w'_j \cdot \texttt{len}(\widehat{\Tb}_j)$
\EndFor

\State \textbf{Return} $\sum_{k=1}^K \frac{R_k - R_{k-1}}{P^K(K \geq k )}$ 

\EndProcedure

\end{algorithmic}
\end{algorithm}

\subsection{Constructing a Sequential Statistical Test via Martingales}
To determine if a provider is misreporting, 
our framework compares the length of the reported sequence of tokens $\widetilde{\Tb}$ as a response to a prompt $Q$ with the estimator of the average length of the sequences used by the LLM to encode the string $\texttt{str}(\widetilde{\Tb})$ using Algorithm~\ref{alg:estimator}. 
More concretely, 
our framework computes the quantity
\begin{equation}\label{eq:e-variable-definition}
        E = \textnormal{\texttt{len}}(\widetilde{\Tb}) - 
        \texttt{EstimateLength}(Q,\texttt{str}(\widetilde{\Tb}),P^\Mcal,P^k),
\end{equation}
which represents the (statistical) evidence in favor of rejecting the hypothesis $H_0$, \ie, flagging the provider as unfaithful. 
%
As the sequential interaction between the auditor and the provider unfolds, our framework aggregates the observed evidence $E$ against $H_0$ at each time step using a (stochastic) process $M$ defined as follows:
\begin{equation}\label{eq:martignale-definition}
    M_i =
        \begin{dcases}
            1 & i=0\\
            M_{i-1} \cdot  (1+\lambda_i \cdot E_i) & i \geq 1,
        \end{dcases}
\end{equation}
where each $E_i$ is defined as in Eq.~\ref{eq:e-variable-definition} for the respective prompt $Q_i$ and reported sequence $\widetilde{\Tb}_i$ at time $i$, 
and $\lambda_i \geq 0$ is a given (potentially time-varying) parameter that weighs the observed evidence.  
%
Here, if one chooses $\lambda_i \approx 0$, the process $M$ stays constant, ignoring any evidence,  and, if one chooses a high value of $\lambda_i$, the process $M$ is very sensitive to any evidence.
Later, we will show that, while in principle high values of $\lambda_i$ are desirable to quickly detect misreporting, the auditor may also risk incorrectly flagging a truthful provider as suspicious, and we will discuss a prescription to address this trade-off.

Leveraging the process $M$, our framework 
flags the provider as unfaithful as soon as the aggregated evidence exceeds a predefined threshold.
More formally, it  rejects $H_0$ using the following (sequential) statistical test:
\begin{equation}\label{eq:phi-test}
    \phi_\alpha = \mathds{1}\left\{ \exists i\in\mathbb{N}\, \colon\, M_i > \frac{1}{\alpha} \right\},
\end{equation}
where $\alpha \in (0,1)$ and the threshold $1/\alpha$ controls how much evidence needs to be gathered by the 
framework to conclude that the provider is (mis-)reporting tokenizations.
Algorithm~\ref{alg:test} summarizes the overall procedure followed by our framework.

\setlength{\textfloatsep}{10pt}
\begin{algorithm}[t]
\caption{It audits a provider for token misreporting.}
\label{alg:test}
\begin{algorithmic}[1]

\State \textbf{Input} Distribution of prompts $P^Q$, next-token probabilities $P^\Mcal$, sequence $\lambda_i\in\mathbb{R}_+$, bound on the false positive rate $\alpha\in(0,1)$, distribution $P^K$.

\State \textbf{Initialize} $M_0\gets 1$, $\texttt{misreport}\gets \text{False}$ 

\For{$i = 1,2,\dots$}
    \State Sample $q_i \sim P^{Q}$ 
    \State Query the provider with $q_i$ to obtain $\tilde{\tb}_i$ 
    \State $s_i\gets\texttt{str}(\tilde{\tb}_i)$
        
        
    \vspace{1mm}\LineComment{Estimate the avg. length of such tokenizations}
    \State $l_i\gets\texttt{EstimateLength}(q_i,s_i,P^\Mcal,P^K)$
    

    \vspace{1mm}\LineComment{Compute the evidence for misreporting}
    \State $E_i \gets \texttt{len}(\tilde{\tb}_i) 
     -l_i    $ 
     

    \vspace{1mm}\LineComment{Aggregate all evidence so far}
    \State $M_i \gets M_{i-1}\cdot \left(1 + \lambda_i \cdot E_i \right)$ 
    

    \vspace{1mm}\LineComment{If the aggregated evidence is sufficient, flag the provider as misreporting}
    \If{$M_i > 1/\alpha$}   
        \State $\texttt{misreport}\gets \text{True}$
        
        \State \Return $ \texttt{misreport}$ 
    \EndIf
    
\EndFor

\end{algorithmic}
\end{algorithm}

%
Importantly, since the function \texttt{EstimateLength} returns unbiased estimates of the average length of the token sequences used by the LLM $\Mcal$ to encode a given string $\sbb$, 
we can first show that, if the provider is faithful, the process $M$ is a martingale, as formalized by the following proposition:
\begin{proposition}\label{prop:martingale}
    Assume the provider is faithful and implements the reporting policy $\pi_0$. Then, for any sequence $\lambda_i$, the process $M$ defined by Eq.~\ref{eq:martignale-definition} is a martingale. 
    That is, for each time step $i$, it holds that:
    \begin{equation}
        \EE_{H_0}\left[ M_i \given M_{i-1} \right]\\ = M_{i-1}
    \end{equation}
\end{proposition}
In words, the above result shows that, on average, if the provider is faithful, the evidence for misreporting given by the process $M$ does not increase over time. 
Consequently, one may expect the test $\phi_\alpha$ defined by Eq.~\ref{eq:phi-test} not to flag a faithful provider as unfaithful.
The next theorem formalizes this expectation by giving sufficient conditions under which the test $\phi_\alpha$ provably controls the probability of falsely flagging a faithful provider, that is, the false positive rate when testing $H_0$ against $H_1$:
\begin{theorem}\label{thm:validity}
    %
    If $1 + \lambda_i \cdot E_i \geq 0$ for all $i \geq 1$ under $H_0$, then, the false positive rate $P_{H_0}\left( \phi_\alpha =1 \right) \leq \alpha$,
    where the probability $P_{H_0}$ is taken across all random variables appearing in Algorithm~\ref{alg:test}.
\end{theorem}

%
%
%
%

%
Intuitively, the above theorem tells us that, as long as the weights $\lambda_i$ are sufficiently small relative to the negative values of $E_i$ originating from sampled tokenizations $\tb$ with below-average length, then the likelihood that the process $M$ exceeds the threshold $1/\alpha$ remains below $\alpha$.
However, it does not rule out the possibility that an unfaithful provider goes undetected. 
In what follows, we provide sufficient conditions under which the number of time steps $\tau=\inf \{i \,\colon\, M_i >1/\alpha \}$ needed for our test $\phi_{\alpha}$ to flag an unfaithful provider is finite:
\begin{theorem}\label{thm:consistency}
    The (sequential) statistical test $\phi_{\alpha}$ satisfies the following:
    \begin{enumerate}[label=\roman*)]
        \item \textbf{Decreasing} $\lambda_i$\textbf{:} Let $\lambda_i = \lambda_0/i > 0$ for all $i \geq 1$. If $1 + \lambda_0 \cdot E > 0$ under $H_0$, then, it holds that 
        \begin{equation*}
            P_{H_1}(\tau < \infty)=1\,\,\,\text{and}\,\,\,\EE_{H_1}\left[\tau \right] < \infty.    
        \end{equation*}
        \item \textbf{Constant} $\lambda_i$\textbf{:} Let $\lambda_i = \lambda_0 > 0$ for all $i \geq 1$. If
        %
            $1+\lambda_0\cdot E \in (B_-, B_+)$ with $B_- >0$ under $H_0$ 
        %
        and
        \begin{equation*}
            \log(1+\lambda_0\cdot \Ical(\pi)) > \textnormal{Var}(E)\cdot\frac{\lambda_0^2}{2(\mathrm{B}_{-})^2} \,\, \text{under} \,\, H_1,
        \end{equation*}
        then, it holds that $P_{H_1}(\tau < \infty)=1$ and
            \begin{equation*}
                            \EE_{H_1}\left[\tau \right] \leq \frac{\log 1/\alpha + \log \mathrm{B}_{+}}{\log (1+\lambda_0\cdot \Ical(\pi)) -  \textnormal{Var}(E)\cdot\frac{\lambda_0^2}{2(\mathrm{B}_{-})^2}}.
             \end{equation*}
        
    \end{enumerate}
\end{theorem}
The above theorem gives sufficient conditions for the execution of Algorithm~\ref{alg:test} to terminate.
Loosely speaking, at each time step $i$, the average evidence that the framework observes is $\EE_{H_1}[1+\lambda_i\cdot E_i]=1+\lambda_i\cdot \Ical(\pi)$ and, if the provider indeed misreports and $\lambda_i$ is sufficiently small, the process $M_i$ grows exponentially fast and will eventually surpass the threshold $1/\alpha$.
Importantly, Theorem~\ref{thm:consistency} makes no assumption on the specific form of the provider'{}s reporting policy $\pi$, and it ensures our auditing framework enjoys worst-case guarantees based on the magnitude $\Ical(\pi)$ of misreporting.

In the next section, we validate our auditing framework by auditing both providers that faithfully report tokenizations and providers that misreport with positive intensity.

\xhdr{Remark} We have considered two simple choices for the sequence $\lambda_i$, namely constant and proportional to $1/i$, however, one could, in principle, construct a more sophisticated weighting sequence to lower the number of steps $\EE_{H_1}[\tau]$ needed to detect an unfaithful provider. In particular, to detect an unfaithful provider in the lowest number of steps, the auditor could consider the weight $\lambda_i$, which, in hindsight, would have led to the higher (logarithmic) growth for the process $\Mcal$, \ie,

\begin{equation}
    \lambda_i = \min\left( \argmax_{\lambda\geq 0} \frac{1}{i-1} \sum_{s=1}^{i-1} \log\left( 1 + \lambda \cdot E_s\right), \lambda_{-} \right),
\end{equation}
where $\lambda_{-}$ is the largest weight such that $1+\lambda_{-}\cdot E \geq 0$, which is required for Theorem~\ref{thm:validity} to hold. The above construction can be shown to be asymptotically optimal in terms of detection time~\citep{waudbysmith2025universallogoptimalitygeneralclasses}. 
Unfortunately, obtaining meaningful bounds on $\EE_{H_1}[\tau]$ that generalize Theorem~\ref{thm:consistency} becomes challenging. 

\begin{figure*}[t]
    \subfloat[\texttt{Llama-3.2-1B-Instruct}]{
    \includegraphics[width=0.33\linewidth, clip, trim={0 0 0 0.5cm}]{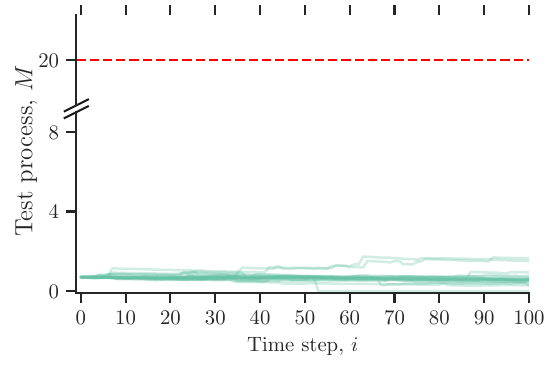}
    }
    \hspace{-5mm}
    \subfloat[\texttt{Ministral-8B-Instruct-2410}]{
    \includegraphics[width=0.33\linewidth, clip, trim={0 0 0 0.5cm}]{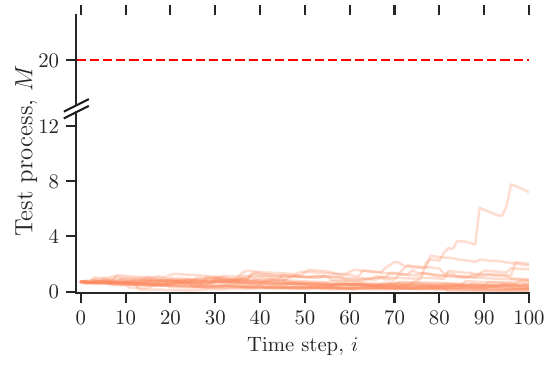}
    }
    \hspace{-5mm}
    \subfloat[\texttt{Gemma-3-1B-It}]{
    \includegraphics[width=0.33\linewidth, clip, trim={0 0 0 0.5cm}]{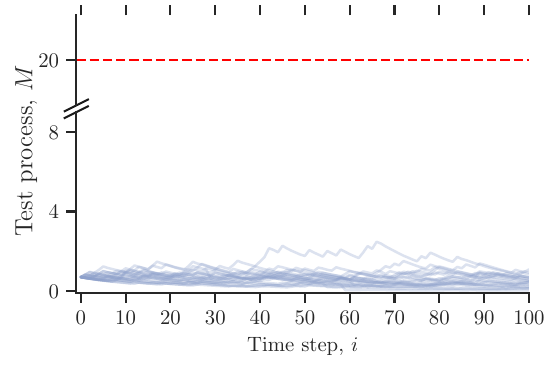}
    }
    \hspace{-5mm}
    \caption{\textbf{Auditing faithful providers.} The panels show realizations of the test process $M$ for three (simulated) faithful providers, each serving a different large language model. 
    In each realization, 
    we sequentially query the provider using prompts picked uniformly at random from the LMSYS Chatbot Arena dataset,
    and compute $M_i$ using Eq.~\ref{eq:martignale-definition} with $\lambda=0.07, 0.13$ and $0.19$, respectively.
    In all panels, the dashed line illustrates the threshold $1/\alpha$ needed to flag a provider and,
    for clarity, we display $30$ realizations randomly sampled from a total of $150$.
    %
    Moreover, we set the false positive rate 
    bound to $\alpha = 0.05$
    and the temperature of the models to $1$. Refer to Appendix~\ref{app:additional-experimental-results} for qualitatively similar results using other temperature values.
    %
    }
    \label{fig:audit-faithful}
\end{figure*}

\section{Experiments}
\label{sec:experiments}
In this section, we use our framework to audit several (simulated) providers who serve large language models from the \texttt{Llama}, \texttt{Gemma} and \texttt{Ministral} families.
Here, we experiment with both providers who are faithful (and use the faithful reporting policy $\pi_0$), and providers who are unfaithful and use the (mis-)reporting policies introduced by~\citet{velasco2025llmoverchargingyoutokenization}.

\xhdr{Experimental setup}
To instantiate our auditing framework, we sequentially query the (simulated) providers, who serve one model from the above-mentioned families, using prompts picked uniformly at random from a set of $4{,}000$ prompts from the LMSYS Chatbot Arena platform dataset~\citep{zheng2024lmsyschat1mlargescalerealworldllm}. 
Refer to Appendix~\ref{app:experimental-details} for further details regarding the dataset and models used by the (simulated) providers. 
Within our framework, we use a Poisson distribution $P^K$ with parameter $7$ to control the number of samples used by our estimator in Algorithm~\ref{alg:estimator},\footnote{We have experimented with Poisson and geometric distributions and have found that $K\sim\text{Poisson(7)}$ performs the best in achieving low variance in the estimator with a small number of samples. However, note that Proposition~\ref{prop:unbiased-estimator} guarantees that our estimator is unbiased for any distribution $P^K$ with support over $\mathbb{N}$.} 
and set $\lambda_i = \lambda$ to a model specific constant value for all $i \geq 1$, which we determine as follows. 
First, we generate $400$ realizations of the random variable $E$ defined in Eq.~\ref{eq:e-variable-definition} using the faithful reporting policy $\pi_0$ on prompts picked uniformly at random from a held-out set of $400$ prompts from the LMSYS Chatbot Arena platform dataset. 
Then, we compute the largest value of $\lambda$ (namely, $\lambda_{-}$) that ensures that \mbox{$1 + \lambda\cdot E$} is positive for every realization of $E$, and set $\lambda=0.9\lambda_{-}$.
This results in $\lambda=0.07, 0.13$ and $0.19$ for \texttt{Llama-3.2-1B-Instruct}, \texttt{Ministral-8B-Instruct-2410} and \texttt{Gemma-3-1B-It}, respectively.
We empirically verify that this choice of $\lambda$ always leads to positive realizations of the process $M$, as required by Theorem~\ref{thm:validity}.
%
Lastly, we set the false positive rate 
bound $\alpha$ to $0.05$, and terminate each audit 
once the testing process $M_i$ exceeds the threshold $1/\alpha$ 
or the audit performs a maximum of $100$ iterations.

\vspace{1mm}\noindent\textbf{Can our audit falsely flag a faithful provider as unfaithful?} 
To answer this question, we audit several (simulated) providers who serve each a different model and use the faithful reporting policy $\pi_0$.
We repeat each audit $150$ times and, each time, we measure how the test process $M$ evolves over time.
Figure~\ref{fig:audit-faithful} summarizes the results, which show that, 
as expected from the martingale property of $M$ shown in Proposition~\ref{prop:martingale}, the vast majority of realizations of the test process $M$ remain very close to their initial value $M_0=1$, with no apparent drift towards larger positive values. 
Moreover, the results also show that, across all 
(simulated) providers, none of the realizations of the test process $M$ exceed the threshold of $1/\alpha$.
This suggests that, although our audit is designed to have a false positive rate at most $\alpha=0.05$ based on Theorem~\ref{thm:validity}, the probability of falsely flagging a faithful provider as unfaithful is negligible in practice.\footnote{This may be due to the fact that Theorem~\ref{thm:validity} relies on Ville's inequality~\citep{ville1939étude}, which leads to conservative upper bounds.}

\begin{figure*}[t]
    \vspace{-1cm}
    \centering

    \subfloat[Random policies]{
    \includegraphics[width=0.46\linewidth]{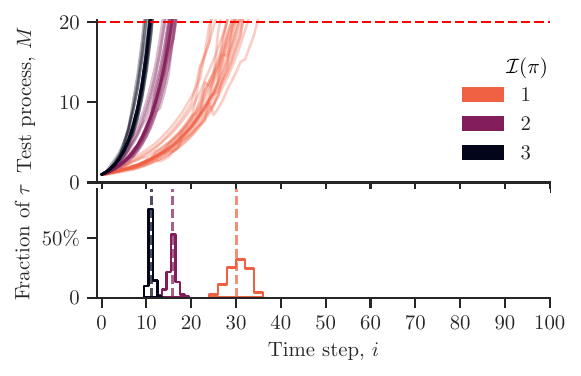}
    }
    \hspace{-5mm}
    \subfloat[Heuristic policies]{
    \includegraphics[width=0.46\linewidth]{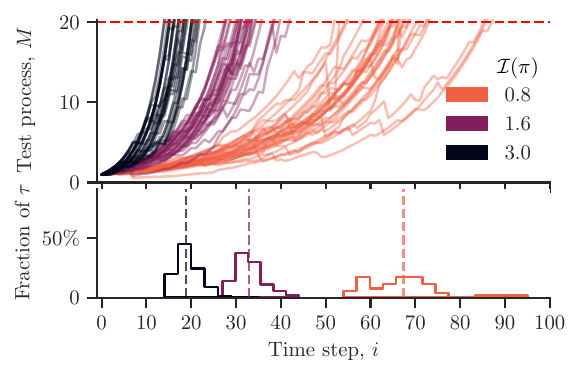}
    }
    \hspace{-5mm}
    \caption{\textbf{Auditing an unfaithful provider who serves \texttt{Llama-3.2-1B-Instruct}.}
    The two panels show realizations of the test process $M$ (top)
    and the distribution of detection times $\tau=\inf \{i\, \colon\, M_i >1/\alpha \}$ (bottom)
    when the provider uses, respectively, random and heuristic policies $\pi$ of varying intensity $\Ical(\pi)$.
    For the heuristic policy, we report an estimate of the intensity by averaging the number of additional tokens that it reports.
    In each realization, we sequentially query the provider using prompts picked uniformly at random from the LMSYS Chatbot Arena dataset, and compute $M_i$ using Eq.~\ref{eq:martignale-definition} with $\lambda=0.07$.
    In each panel, the three different intensity values correspond to policies $\pi$ parameterized by $m=1,2,3$, with higher values of $m$ leading to higher (darker) intensities and, for each $m$, we show $30$ realizations.
    In all panels, we set the 
    false positive rate bound to $\alpha = 0.05$ and the temperature of the models to $1$. Refer to Appendix~\ref{app:additional-experimental-results} for qualitatively similar results using other models and temperature values.
    }
    \label{fig:test-all}
\end{figure*}

\vspace{1mm}\noindent\textbf{Can our audit successfully detect an unfaithful provider?}
To answer this question, we audit several (simulated) providers who serve each a different model and use two different types of (mis-)reporting policies~\citep{velasco2025llmoverchargingyoutokenization}.
These policies construct a tokenization $\tilde{\tb}$ of the string $\texttt{str}(\tb)$ by iteratively selecting $m$ tokens and splitting them into two separate tokens.
We briefly describe the two types of misreporting policies below and provide their full description in Appendix~\ref{app:policies}: 

\begin{itemize}
    \item \textbf{Random:} 
        For each generated output, it iteratively selects a random token to modify and splits it into a random pair of tokens (if any). It terminates once it has performed $m$ splits and reports the modified sequence $\tilde{\tb}$.
        As a consequence, the misreporting intensity $\Ical(\pi)$ for a policy $\pi$ of this type is (approximately) $m$.
    \item \textbf{Heuristic:} 
        For each generated output, it prioritizes reported tokenizations that are not very unlikely to be generated.
        To this end, in each iteration, it selects the token with the highest index in the vocabulary and splits it into a pair of tokens with the highest minimum index. Once it has performed $m$ successful splits resulting in an (intermediate) modified token sequence $\tb'$, the policy computes the next-token probabilities given by the model $\Mcal$ at each point in $\tb'$ and verifies if each token is in its respective top-$p$ set~\citep{Holtzman2020The}.
        If this condition is satisfied, the policy reports $\tilde{\tb}=\tb'$ to the user; otherwise, it falls back to $\tilde{\tb}=\tb$. 
        Consequently, the misreporting intensity $\Ical(\pi)$ for a policy $\pi$ of this type is at most $m$, depending on the fraction of modified tokenizations $\tb'$ that pass the verification step, and it can be estimated by computing the average number of additional tokens reported.
\end{itemize}
Figure~\ref{fig:test-all} summarizes the results for an unfaithful provider who serves a model from the \texttt{Llama} family.
Refer to Appendix~\ref{app:additional-experimental-results} for qualitatively similar results for unfaithful providers who serve other models.
We find that, in all cases, our auditing framework succeeds in detecting token misreporting (\ie, the test process $M$ crosses the threshold $1/\alpha$).
Importantly, the results also show that our framework detects token misreporting after observing fewer than $\sim$$70$ reported outputs.
Moreover, consistent with Theorem~\ref{thm:consistency}, we observe that the average number of necessary reported outputs to detect token misreporting decreases rapidly as the misreporting intensity $\Ical(\pi)$ increases. 
However, we also find that auditing an unfaithful provider who uses the heuristic policy rather than the random policy generally leads to a higher variance in the number of queries needed to flag them, which, in some cases, can increase the time and resources needed to detect misreporting.

\section{Discussion and Limitations}
\label{sec:discussion}
In this section, we highlight several limitations of our work and discuss avenues for future research as well as its broader impact.

\xhdr{Model access}
%
%
Our auditing framework's strong theoretical guarantees (\ie, Theorems~\ref{thm:validity} and~\ref{thm:consistency}) require sandboxed or regulatory access to the language model served by the provider.
We believe that some form of model access is necessary for any effective auditing method 
%
since detecting manipulations in reported token sequences fundamentally requires a baseline reflecting the behavior of the unmanipulated model; in our framework, the \texttt{EstimateLength} estimator provides precisely such a baseline.
Nevertheless, in future work, it would be very interesting to develop auditing techniques relying on weaker forms of model access. For example, one promising direction is to investigate whether access limited to the model'{}s tokenizer could still enable meaningful audits.
In this context, it is also important to note that, in high-stakes settings, regulatory regimes such as the EU AI Act, Article 74(13) already foresee that trusted third-party auditors, whether public or private, may be granted privileged access to proprietary models. 

\xhdr{Reporting policies}
In our theoretical analysis, we have assumed that the reporting policy is static because this allows us to obtain closed-form bounds for the number of steps $\EE_{H_1}[\tau]$ required to detect misreporting in Theorem~\ref{thm:consistency}. 
However, our auditing framework can, in principle, be instantiated even if the provider varies their reporting policy over time. This is because the validity of Theorem~\ref{thm:consistency} (\ie, the condition $1 + \lambda_i \cdot E_i \geq 0 $ under $H_0$) does not require any assumption on the provider'{}s reporting policy, and because the estimator in Algorithm~\ref{alg:estimator} remains unbiased at each time step $i$, regardless of how the policy may change at the next step $i+1$. 
That said, a non-static policy would require a more refined theoretical analysis to obtain guarantees on $\EE_{H_1}[\tau]$. Intuitively, if the provider uses a time-varying misreporting policy $\pi_i$ at each time step $i$, then, on expectation, the test process $M$ is multiplied at each step by the quantity $1+\lambda_i\cdot \Ical(\pi_i)$, where the misreporting intensity $\Ical(\pi_i)$ is no longer constant. Thus, if the intensities decay over time (\eg, as the provider becomes more cautious), detection is expected to become increasingly difficult---though misreporting also becomes correspondingly less harmful to the user.

Further, we would like to emphasize that our auditing framework can use historical billing data collected offline, \ie, lines 4 and 5 in Algorithm~\ref{alg:test} can take place before the provider learns they will be audited. 
For example, the reported tokenizations may come from users who suspect misreporting and save the tokenizations reported to them. 
As a consequence, even if the provider has stopped misreporting at the time the audit is taking place, they may be unable to evade detection if they have misreported in the past.

\xhdr{Evaluation}
We have conducted experiments with state-of-the-art open-weight LLMs from the \texttt{Llama}, \texttt{Gemma} and \texttt{Ministral} families. However, it would be interesting to evaluate of our auditing framework using proprietary LLMs, as well as conducting evaluations with real providers. 
Such evaluations are out of scope of the current work since they would come with additional technical and regulatory challenges; however, we believe they may prove useful at building trust between users and providers in LLM-as-a-service.
Moreover, we have validated our framework against the set of \mbox{(mis-)reporting} policies recently introduced by~\cite{velasco2025llmoverchargingyoutokenization}. 
In practice, however, providers may employ other policies.
Documenting and systematizing such policies---similarly to how the jailbreaking literature collects adversarial prompts~\citep{rao2024trickingllmsdisobedienceformalizing, shen2024donowcharacterizingevaluating}---would provide a valuable benchmark for future auditing frameworks.

\xhdr{Broader impact}
As the multi-billion-dollar market of LLM-as-a-service keeps growing, it is increasingly critical to develop statistical techniques for algorithmic auditing, especially since the provider'{}s incentives under the current pay-per-token pricing model are misaligned with those of end-users~\citep{velasco2025llmoverchargingyoutokenization}. 
Our work enables a trusted third-party auditor with access to the LLMs served by providers to detect token misreporting 
regardless of the (mis)-reporting policy used by unfaithful providers, however, the potential impact of our framework will (partially) depend on the timely development of regulatory frameworks for LLMs and, more broadly, generative AI. 

\section{Conclusions}
\label{sec:conclusions}
In this work, we have introduced a first-of-a-kind auditing framework that enables a third-party trusted auditor with access to the LLM served by a provider to detect token misreporting.
Along the way, we developed an unbiased estimator of the average length of token sequences that an LLM uses to encode a string, and we established conditions under which our framework is guaranteed to detect an unfaithful provider---regardless of their (mis)reporting policy---while avoiding falsely flagging faithful providers with high probability.

Furthermore, we have empirically validated our au\-di\-ting framework across a broad range of misreporting policies and LLM families, and showed that an auditor needs to query an unfaithful provider only about $\sim$$70$ times to reliably flag them.
More broadly, we hope that our work will raise awareness of the urgent need to develop auditing tools that discourage LLM providers from engaging in unfaithful practices and protect users, who are vulnerable under the current pay-per-token pricing model.

\vspace{2mm}
\xhdr{Acknowledgements} 
Gomez-Rodriguez acknowledges support from the European Research Council (ERC) under the European Union'{}s Horizon 2020 research and innovation programme (grant agreement No. 945719 and 101169607).
Tsirtsis acknowledges supports from the Alexander von Humboldt Foundation in the framework of the Alexander von Humboldt Professorship (Humboldt Professor of Technology and Regulation awarded to Sandra Wachter) endowed by the Federal Ministry of Education and Research via the Hasso Plattner Institute.

{ 
\small
\bibliographystyle{plainnat}
\bibliography{token-audit}
}

\clearpage
\newpage

\appendix

\section{Additional Experimental Details}\label{app:experimental-details}

Here, we provide additional details on the experimental setup, including the hardware used, the dataset and models used, as well as details on the generation process.\footnote{All code is publicly available at \url{https://github.com/Human-Centric-Machine-Learning/token-audit}.}

\xhdr{Hardware setup} Our experiments are executed on a compute server equipped with 2 $\times$ Intel Xeon Gold 5317 CPU, $1{,}024$ GB main memory, and $2$ $\times$ A100 Nvidia Tesla GPU ($80$ GB, Ampere Architecture). In each experiment, a single Nvidia A100 GPU is used.

\xhdr{Generation details} We use \texttt{Python 3.11} and the \texttt{transformers} library\footnote{\url{https://github.com/huggingface/transformers}} as the API to run the models. In all results presented in Section~\ref{sec:experiments}, we set the temperatures of the models to $1$, and all outputs used are generated with no top-$p$ sampling. In Appendix~\ref{app:additional-experimental-results}, we present additional results with other temperature values. 
We instruct LLMs to generate responses to the LMSYS Chatbot Arena dataset prompts by using the following system prompt for all models:

\begin{table}[h]
\centering
\begin{tcolorbox}[
    colframe=white,      
    colback=gray!14,     
    boxrule=0.5mm,       
    arc=4mm,             
    left=3mm,            
    right=3mm,           
    top=3mm,             
    bottom=3mm           
]
\begin{tabular}{ m{0.97\textwidth} }
    \rowcolor{gray!14}
    \textbf{System:} You are a helpful assistant. Answer briefly and to the point.
\end{tabular}
\end{tcolorbox}
\vspace{-12mm}
\caption*{}
\label{app:sys_prompt_mmlu}
\end{table}
When implementing Algorithm~\ref{alg:heuristic} to construct the reported tokenizations, we use the specified top-$p$ value to verify if the sequence $\tb'$ satisfies the condition $t'_i \in \Vcal_p(\tb'_{\leq i-1}) \,\,\forall i \in [ \texttt{len}\left( \tb'\right)]$, where $\Vcal_p(\tb'_{\leq i-1})$ is the smallest subset of $\Vcal$ whose cumulative next-token probability is at least $p\in(0,1)$.

\xhdr{Datasets}
For the results presented in all figures, we generated model responses to prompts obtained from the LMSYS-Chat-1M dataset~\citep{zheng2024lmsyschat1mlargescalerealworldllm}. We use the LMSYS-Chat-1M dataset exclusively to obtain a varied sample of potential user prompts. We filter user prompts to obtain the first 4000 questions that are in English (by using the \texttt{language} keyword) and whose length (in number of characters) is in the range $[20, 100]$, to avoid trivial or overly elaborated prompts. We have repeated our experiments with a different set of 4000 randomly selected prompts from the LMSYS-Chat-1M dataset and have found indistinguishable results.

\xhdr{Models} In our main experiments, we use the model \texttt{Llama-3.2-3B-Instruct} from the \texttt{Llama} family, the model \texttt{Gemma-3-1B-It} from the \texttt{Gemma} family, and \texttt{Ministral-8B\-Instruct-2410}. In Appendix~\ref{app:robust}, we use three additional quantized versions of the above models, namely \texttt{RedHatAI/Llama-3.2-1B-Instruct-FP8}, \texttt{RedHatAI/gemma-3-1b-it-quantized.w8a8} and \texttt{QuantFactory/Ministral-8B-Instruct-2410-GGUF}. The models are obtained from publicly available repositories from \texttt{Hugging Face}\footnote{\url{https://huggingface.co/meta-llama/Llama-3.2-3B-Instruct} \newline \url{https://huggingface.co/meta-llama/Llama-3.2-1B-Instruct}\newline \url{https://huggingface.co/google/gemma-3-1b-it}\newline \url{https://huggingface.co/google/gemma-3-4b-it}
\newline \url{https://huggingface.co/mistralai/Ministral-8B-Instruct-2410}
\newline
\url{https://huggingface.co/RedHatAI/Llama-3.2-1B-Instruct-FP8}
\newline
\url{https://huggingface.co/RedHatAI/gemma-3-1b-it-quantized.w8a8}
\newline
\url{https://huggingface.co/QuantFactory/Ministral-8B-Instruct-2410-GGUF}
}.

\clearpage
\newpage

\section{Additional Experimental Results}\label{app:additional-experimental-results}
In this section, we provide additional experimental results for our auditing framework under different LLMs and temperature parameters.

\subsection{Audit Results for a Faithful Provider}

\begin{figure}[ht]
    \centering

    \subfloat{\parbox[b]{0.05\textwidth}{\centering\rotatebox[origin=c]{90}{\hspace{0.1cm}\scriptsize Temperature $1.0$}}}%
    \subfloat{\includegraphics[width=0.31\textwidth, clip, trim={0 0 0 0.5cm}]{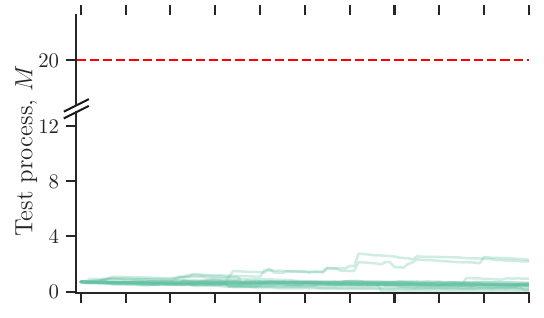}}\hfill
    \subfloat{\includegraphics[width=0.28\textwidth, clip, trim={0 0 0 0.5cm}]{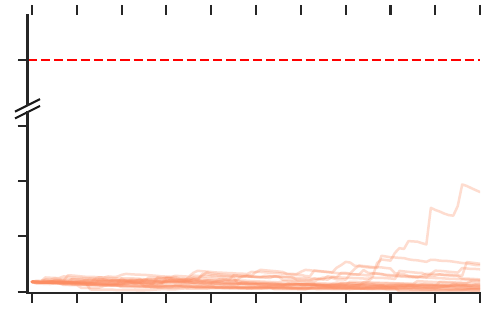}}\hfill
    \subfloat{\includegraphics[width=0.28\textwidth, clip, trim={0 0 0 0.5cm}]{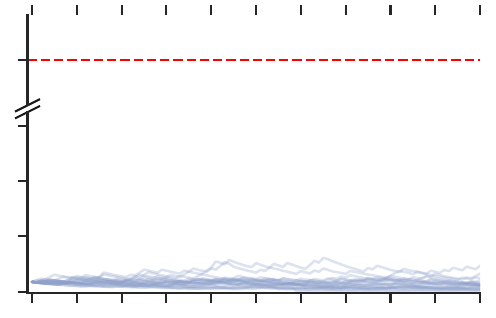}}\\[5pt]
    
    \subfloat{\parbox[b]{0.05\textwidth}{\centering\rotatebox[origin=c]{90}{\hspace{0.1cm}\scriptsize Temperature $1.15$}}}%
    \subfloat{\includegraphics[width=0.31\textwidth, clip, trim={0 0 0 0.5cm}]{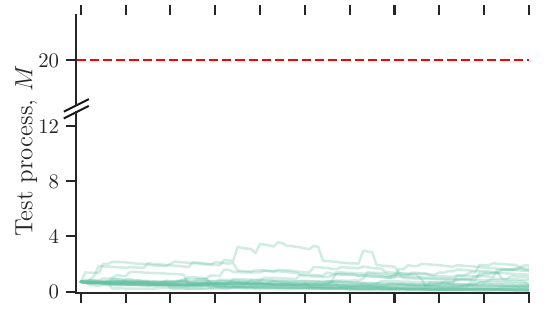}}\hfill
    \subfloat{\includegraphics[width=0.28\textwidth, clip, trim={0 0 0 0.5cm}]{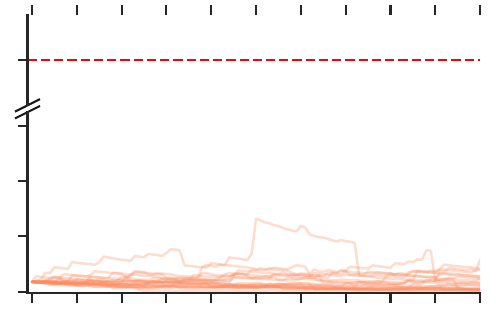}}\hfill
    \subfloat{\includegraphics[width=0.28\textwidth, clip, trim={0 0 0 0.5cm}]{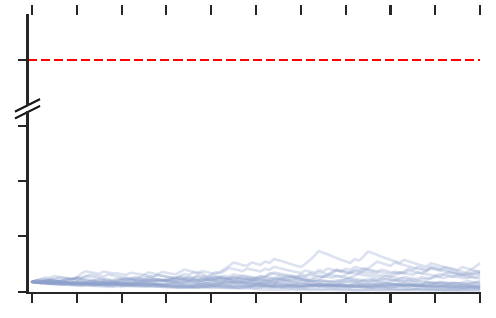}}\\[5pt]
    
    \subfloat{\parbox[b]{0.05\textwidth}{\centering\rotatebox[origin=c]{90}{\hspace{0.6cm}\scriptsize Temperature $1.3$}}}%
    \subfloat{\includegraphics[width=0.31\textwidth, clip, trim={0 0 0 0.5cm}]{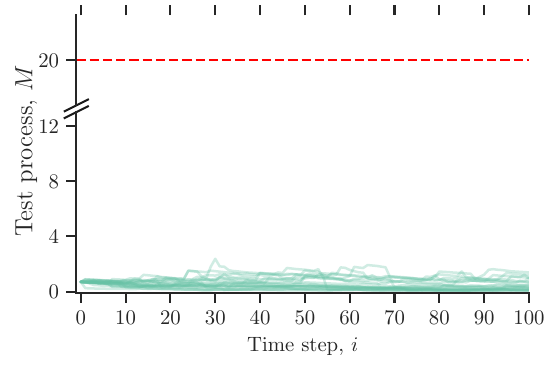}}\hfill
    \subfloat{\includegraphics[width=0.28\textwidth, clip, trim={0 0 0 0.5cm}]{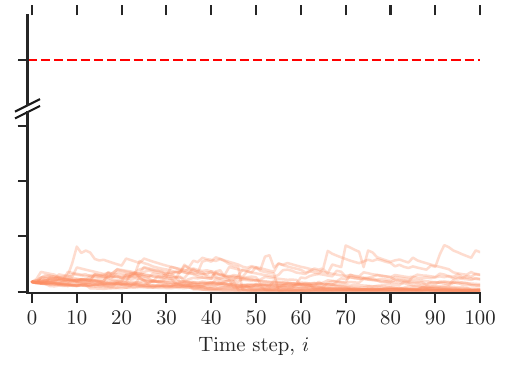}}\hfill
    \subfloat{\includegraphics[width=0.28\textwidth, clip, trim={0 0 0 0.5cm}]{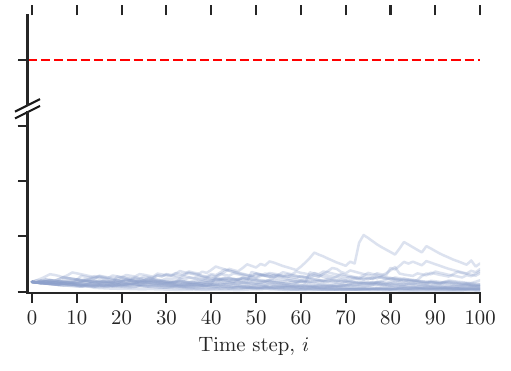}}\\[5pt]

    \subfloat{\hspace{0.05\textwidth}\parbox[b]{0.31\textwidth}{\centering \texttt{Llama-3.2-1B-Instruct}}}\hfill
    \subfloat{\parbox[b]{0.31\textwidth}{\centering \texttt{Ministral-8B-Instruct-2410}}}\hfill
    \subfloat{\parbox[b]{0.31\textwidth}{\centering \texttt{Gemma-3-1B-It}}}

    \caption{\textbf{Auditing faithful providers.} The panels show realizations of the test process $M$ for three (simulated) faithful providers, each serving a different large language model, across different temperature values used during generation and auditing.
    In each realization, 
    we sequentially query the provider using prompts picked uniformly at random from the LMSYS Chatbot Arena dataset,
    and compute $M_i$ using Eq.~\ref{eq:martignale-definition} with $\lambda=0.07, 0.13$ and $0.19$ for temperature $1.0$, $\lambda=0.10, 0.11$ and $0.10$ for temperature $1.0$, and $\lambda=0.10, 0.10$ and $0.19$ for temperature $1.0$, for  \texttt{Llama-3.2-1B-Instruct}, \texttt{Ministral-8B-Instruct-2410} and \texttt{Gemma-3-1B-It},  respectively.
    In all panels, the dashed line illustrates the threshold $1/\alpha$ needed to flag a provider and,
    for clarity, we display $30$ realizations randomly sampled from a total of $150$. Moreover, we set the false positive rate 
    bound to $\alpha = 0.05$.
    }
    \label{fig:app-faithful}
\end{figure}

\clearpage
\newpage

\subsection{Audit Results Using the Random Policies in Algorithm~\ref{alg:random}}

\begin{figure*}[h]
    
    \centering
    \captionsetup[subfigure]{labelformat=empty}
    \subfloat{
    \includegraphics[trim={0 9cm 0 1.5cm},clip,width=0.7\linewidth]{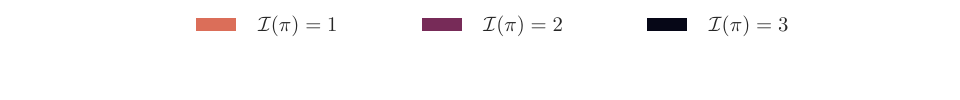}
    }\\[1em]
    \vspace{-5mm}
    \subfloat{\parbox[b]{0.05\textwidth}{\centering\rotatebox[origin=c]{90}{\hspace{0.1cm}\scriptsize Temperature $1.0$}}}%
    \subfloat{\includegraphics[width=0.32\textwidth, clip, trim={0 0 0 0}]{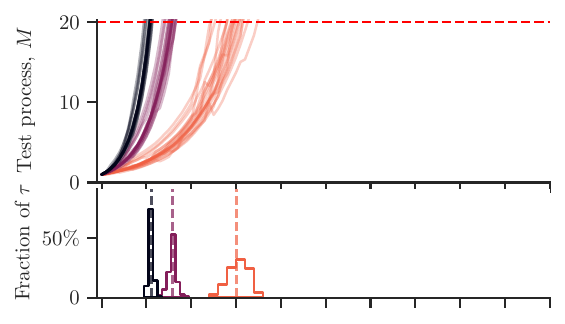}}\hfill
    \subfloat{\includegraphics[width=0.28\textwidth, clip, trim={0 0 0 0}]{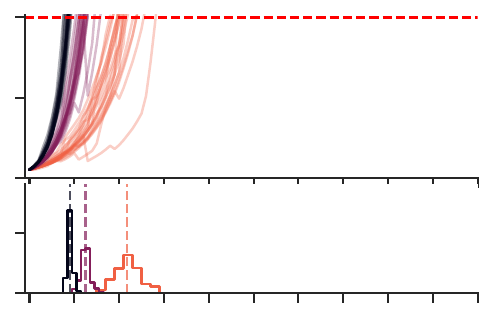}}\hfill
    \subfloat{\includegraphics[width=0.28\textwidth, clip, trim={0 0 0 0}]{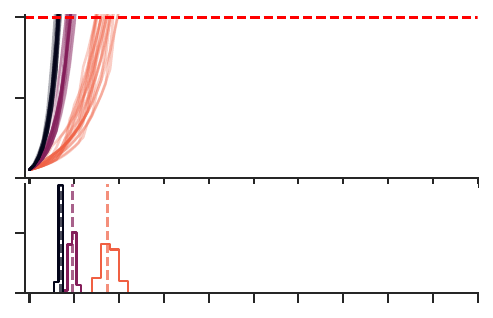}}\\[5pt]
    
    \subfloat{\parbox[b]{0.05\textwidth}{\centering\rotatebox[origin=c]{90}{\hspace{0.1cm}\scriptsize Temperature $1.15$}}}%
    \subfloat{\includegraphics[width=0.32\textwidth, clip, trim={0 0 0 0}]{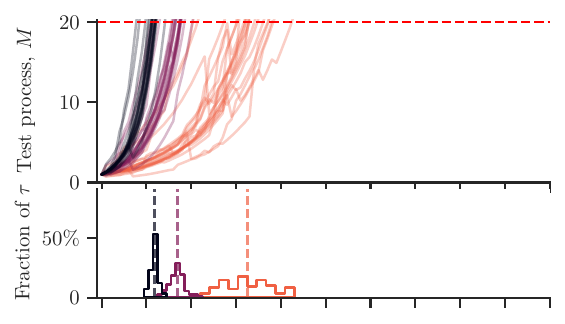}}\hfill
    \subfloat{\includegraphics[width=0.28\textwidth, clip, trim={0 0 0 0}]{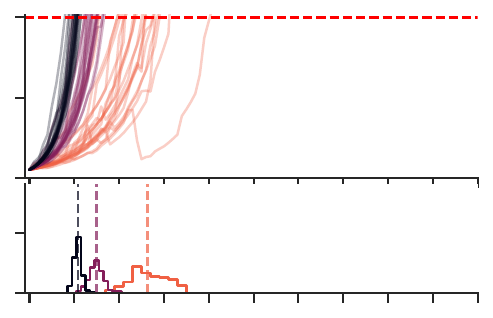}}\hfill
    \subfloat{\includegraphics[width=0.28\textwidth, clip, trim={0 0 0 0}]{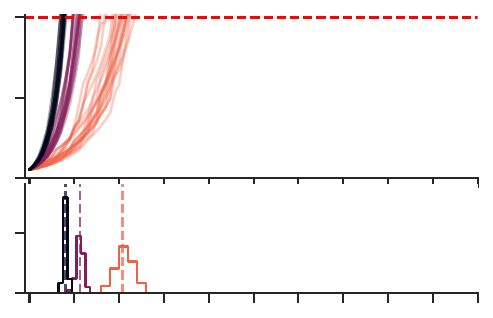}}\\[5pt]
    
    \subfloat{\parbox[b]{0.05\textwidth}{\centering\rotatebox[origin=c]{90}{\hspace{0.6cm}\scriptsize Temperature $1.3$}}}%
    \subfloat{\includegraphics[width=0.32\textwidth]{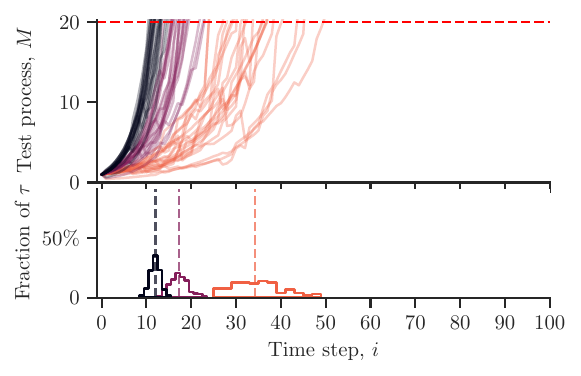}}\hfill
    \subfloat{\includegraphics[width=0.28\textwidth, clip, trim={0 0 0 0}]{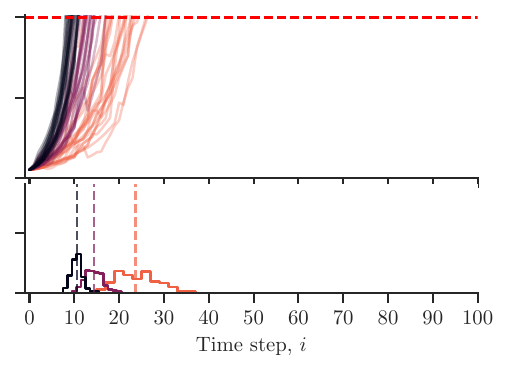}}\hfill
    \subfloat{\includegraphics[width=0.28\textwidth, clip, trim={0 0 0 0}]{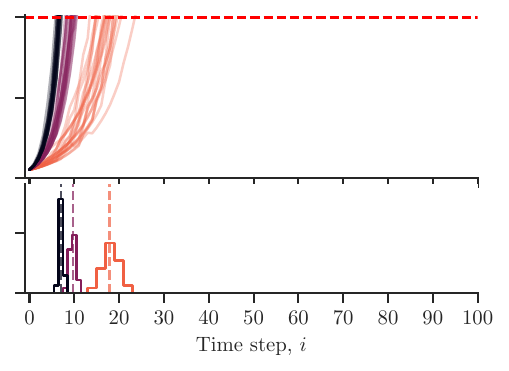}}\\[5pt]

    \subfloat{\hspace{0.05\textwidth}\parbox[b]{0.31\textwidth}{\centering \texttt{Llama-3.2-1B-Instruct}}}\hfill
    \subfloat{\parbox[b]{0.31\textwidth}{\centering \texttt{Ministral-8B-Instruct-2410}}}\hfill
    \subfloat{\parbox[b]{0.31\textwidth}{\centering \texttt{Gemma-3-1B-It}}}
    \hspace{-5mm}
    \caption{\textbf{Auditing an unfaithful provider who misreports using Algorithm~\ref{alg:random}.}
    The panels show realizations of the test process $M$ (top)
    and the distribution of detection times $\tau=\inf \{i\, \colon\, M_i >1/\alpha \}$ (bottom)
    when the provider uses random policies $\pi$ of varying intensity $\Ical(\pi)$, across different models served and temperature values.
    In each realization, we sequentially query the provider using prompts picked uniformly at random from the LMSYS Chatbot Arena dataset, and compute $M_i$ using Eq.~\ref{eq:martignale-definition} with $\lambda=0.07, 0.13$ and $0.19$ for temperature $1.0$, $\lambda=0.10, 0.11$ and $0.10$ for temperature $1.0$, and $\lambda=0.10, 0.10$ and $0.19$ for temperature $1.0$, for  \texttt{Llama-3.2-1B-Instruct}, \texttt{Ministral-8B-Instruct-2410} and \texttt{Gemma-3-1B-It},  respectively.
    In each panel, the three different intensity values correspond to policies $\pi$ parameterized by $m=1,2,3$, with higher values of $m$ leading to higher (darker) intensities, and, for each $m$, we show $30$ realizations.
    In all panels, we set the 
    false positive rate bound to $\alpha = 0.05$.
    }
    \label{fig:app-random}
\end{figure*}

\clearpage
\newpage

\subsection{Audit Results Using the Heuristic Policies in Algorithm~\ref{alg:heuristic}}

\begin{figure*}[h]

    \centering
    
    \subfloat{\parbox[b]{0.05\textwidth}{\centering\rotatebox[origin=c]{90}{\hspace{0.1cm}\scriptsize Temperature $1.0$}}}%
    \subfloat{\includegraphics[width=0.32\textwidth, clip, trim={0 0 0 0}]{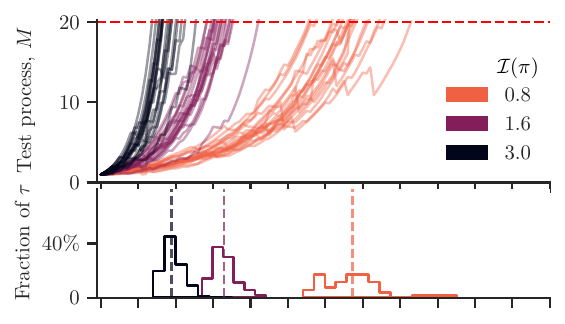}}\hfill
    \subfloat{\includegraphics[width=0.28\textwidth, clip, trim={0 0 0 0}]{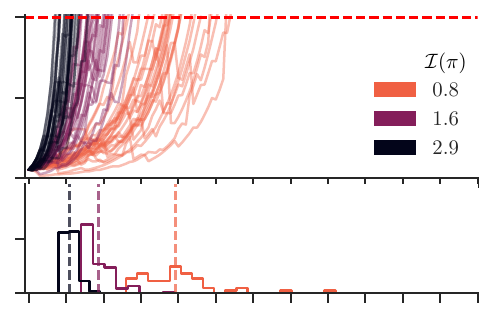}}\hfill
    \subfloat{\includegraphics[width=0.28\textwidth, clip, trim={0 0 0 0}]{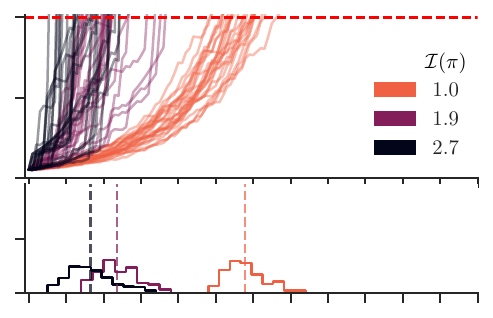}}\\[5pt]
    
    \subfloat{\parbox[b]{0.05\textwidth}{\centering\rotatebox[origin=c]{90}{\hspace{0.1cm}\scriptsize Temperature $1.15$}}}%
    \subfloat{\includegraphics[width=0.32\textwidth, clip, trim={0 0 0 0}]{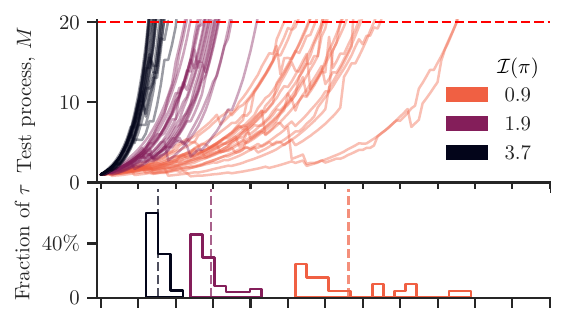}}\hfill
    \subfloat{\includegraphics[width=0.28\textwidth, clip, trim={0 0 0 0}]{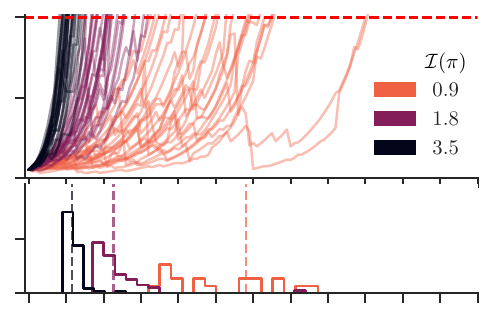}}\hfill
    \subfloat{\includegraphics[width=0.28\textwidth, clip, trim={0 0 0 0}]{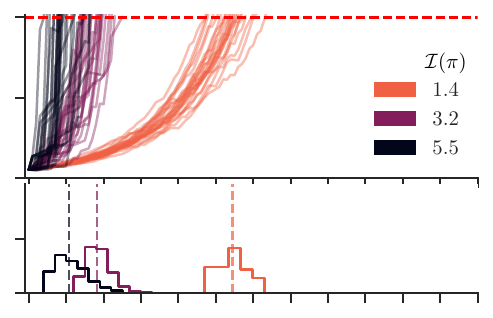}}\\[5pt]
    
    \subfloat{\parbox[b]{0.05\textwidth}{\centering\rotatebox[origin=c]{90}{\hspace{0.6cm}\scriptsize Temperature $1.3$}}}%
    \subfloat{\includegraphics[width=0.32\textwidth]{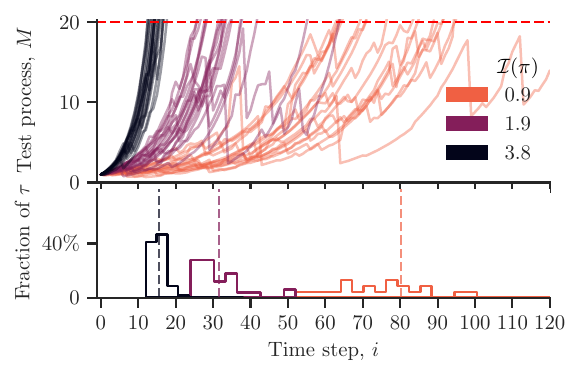}}\hfill
    \subfloat{\includegraphics[width=0.28\textwidth, clip, trim={0 0 0 0}]{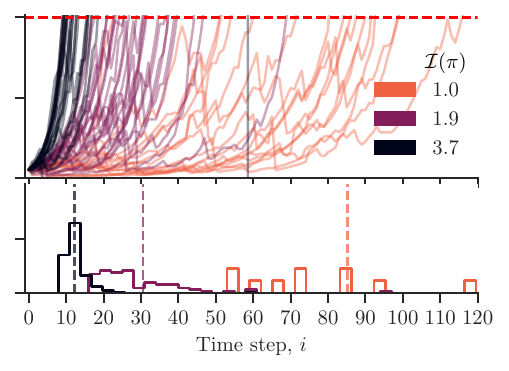}}\hfill
    \subfloat{\includegraphics[width=0.28\textwidth, clip, trim={0 0 0 0}]{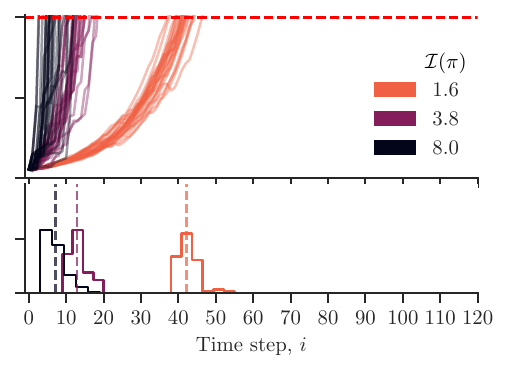}}\\[5pt]
    \subfloat{\hspace{0.05\textwidth}\parbox[b]{0.31\textwidth}{\centering \texttt{Llama-3.2-1B-Instruct}}}\hfill
    \subfloat{\parbox[b]{0.31\textwidth}{\centering \texttt{Ministral-8B-Instruct-2410}}}\hfill
    \subfloat{\parbox[b]{0.31\textwidth}{\centering \texttt{Gemma-3-1B-It}}}
    \hspace{-5mm}
    \caption{\textbf{Auditing an unfaithful provider who misreports using Algorithm~\ref{alg:heuristic}.}
    The panels show realizations of the test process $M$ (top)
    and the distribution of detection times $\tau=\inf \{i\, \colon\, M_i >1/\alpha \}$ (bottom)
    when the provider uses random policies $\pi$ of varying intensity $\Ical(\pi)$, across different models served and temperature values.
    In each realization, we sequentially query the provider using prompts picked uniformly at random from the LMSYS Chatbot Arena dataset, and compute $M_i$ using Eq.~\ref{eq:martignale-definition} with $\lambda=0.07, 0.13$ and $0.19$ for temperature $1.0$, $\lambda=0.10, 0.11$ and $0.10$ for temperature $1.0$, and $\lambda=0.10, 0.10$ and $0.19$ for temperature $1.0$, for  \texttt{Llama-3.2-1B-Instruct}, \texttt{Ministral-8B-Instruct-2410} and \texttt{Gemma-3-1B-It},  respectively.
    In each panel, for clarity, we show $20$ randomly sampled realizations.
    In all panels, we set the 
    false positive rate bound to $\alpha = 0.05$.
    }
    \label{fig:app-heuristic}
\end{figure*}

\clearpage
\newpage

\subsection{Robustness of Algorithm~\ref{alg:test} to Approximate Model Access}\label{app:robust}

\begin{figure}[ht]
    \centering

    \subfloat{\parbox[b]{0.05\textwidth}{\centering\rotatebox[origin=c]{90}{\hspace{1cm}\scriptsize Faithful ($\pi_0$)}}}%
    \subfloat{\includegraphics[width=0.31\textwidth, clip, trim={0 0 0 0.5cm}]{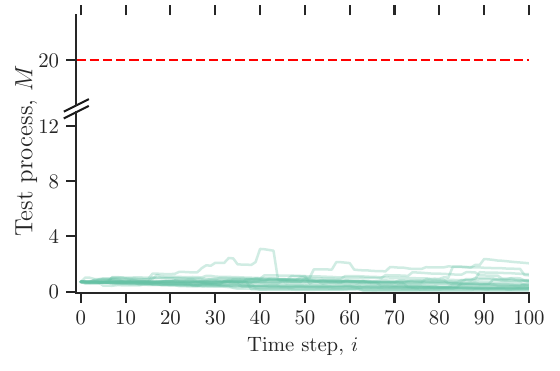}}\hfill
    \subfloat{\includegraphics[width=0.31\textwidth, clip, trim={0 0 0 0.5cm}]{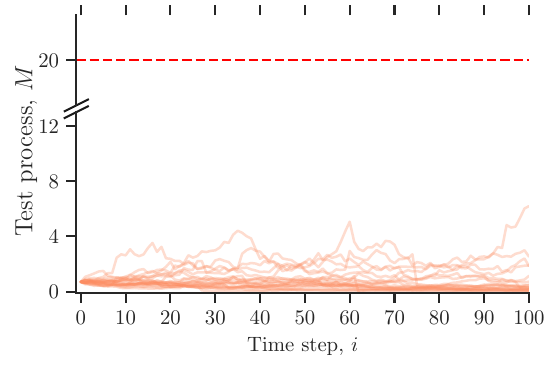}}\hfill
    \subfloat{\includegraphics[width=0.31\textwidth, clip, trim={0 0 0 0.5cm}]{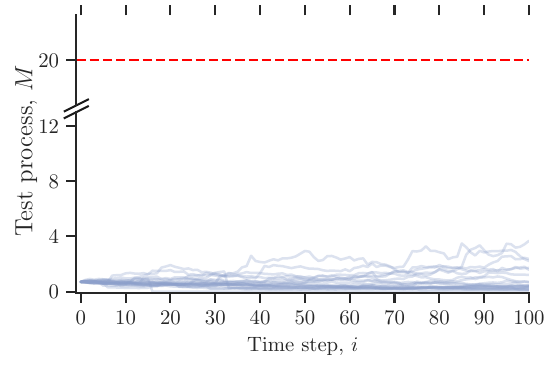}}\\[5pt]
    
    \subfloat{\parbox[b]{0.05\textwidth}{\centering\rotatebox[origin=c]{90}{\hspace{0.7cm}\scriptsize Random (Algorithm~\ref{alg:random})}}}%
    \subfloat{\includegraphics[width=0.31\textwidth, clip, trim={0 0 0 0cm}]{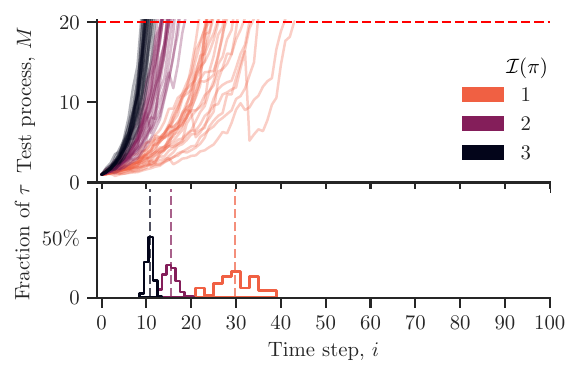}}\hfill
    \subfloat{\includegraphics[width=0.31\textwidth, clip, trim={0 0 0 0cm}]{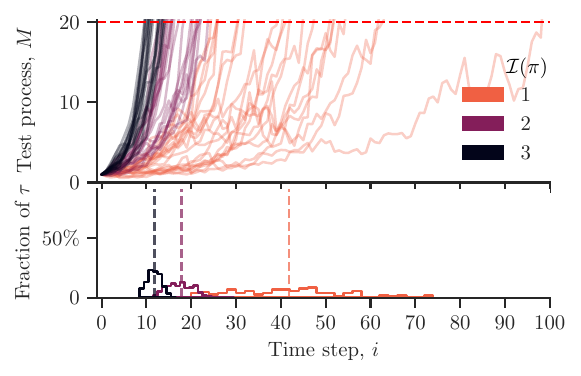}}\hfill
    \subfloat{\includegraphics[width=0.31\textwidth, clip, trim={0 0 0 0cm}]{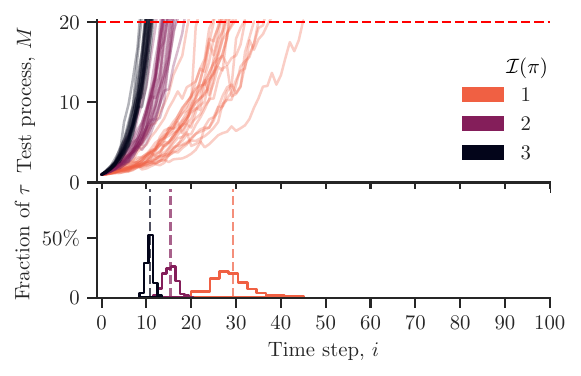}}\\[5pt]
    
    \subfloat{\parbox[b]{0.05\textwidth}{\centering\rotatebox[origin=c]{90}{\hspace{0.6cm}\scriptsize Heuristic (Algorithm~\ref{alg:heuristic})}}}%
    \subfloat{\includegraphics[width=0.31\textwidth, clip, trim={0 0 0 0cm}]{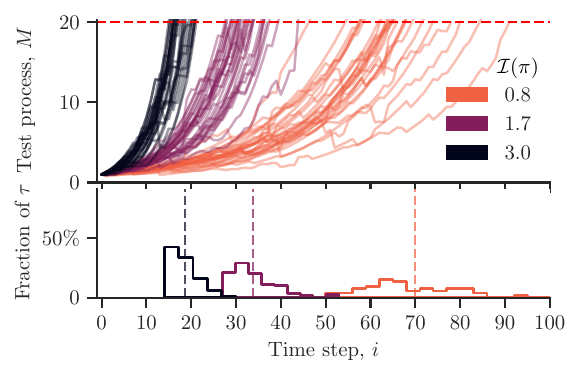}}\hfill
    \subfloat{\includegraphics[width=0.31\textwidth, clip, trim={0 0 0 0cm}]{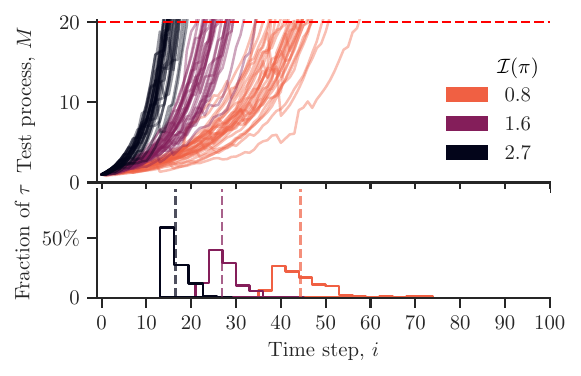}}\hfill
    \subfloat{\includegraphics[width=0.31\textwidth, clip, trim={0 0 0 0cm}]{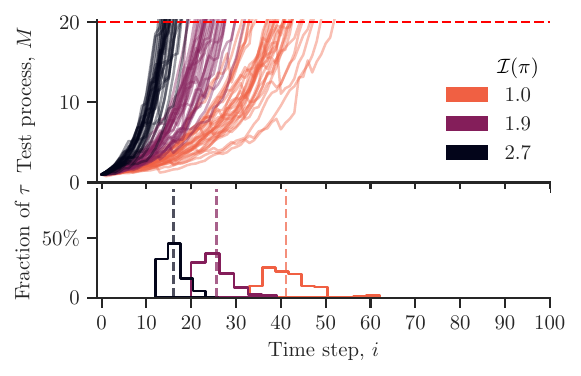}}\\[5pt]

    \subfloat{\hspace{0.05\textwidth}\parbox[b]{0.31\textwidth}{\centering \scriptsize Auditor: \texttt{Llama-3.2-1B-Instruct} \\ Provider: \texttt{Llama-3.2-1B-Instruct-FP8}}}\hfill
    \subfloat{\parbox[b]{0.31\textwidth}{\centering\scriptsize Auditor: \texttt{Ministral-8B-Instruct-2410} \\ Provider: \texttt{Ministral-8B-Instruct-2410-GGUF}}}\hfill
    \subfloat{\parbox[b]{0.31\textwidth}{\centering\scriptsize Auditor: \texttt{Gemma-3-1B-It}\\Provider: \texttt{Gemma-3-1B-It-quantized.w8a8}}}

    \caption{\textbf{Auditing providers with approximate model access.} The panels show realizations of the test process $M$ for simulated providers who use quantized versions of the LLMs they serve and report tokenizations using the faithful reporting policy $(\pi_0)$, the random misreporting policy in Algorithm~\ref{alg:random}, and the heuristic misreporting policy in Algorithm~\ref{alg:heuristic}.
    In each realization, 
    we sequentially query the provider using prompts picked uniformly at random from the LMSYS Chatbot Arena dataset,
    and compute $M_i$ using Eq.~\ref{eq:martignale-definition} with $\lambda=0.08, 0.13$ and $0.18$ for  \texttt{Llama-3.2-1B-Instruct}, \texttt{Ministral-8B-Instruct-2410} and \texttt{Gemma-3-1B-It},  respectively.
    In all panels, the dashed line illustrates the threshold $1/\alpha$ needed to flag a provider and,
    for clarity, we display $30$ realizations randomly sampled from a total of $150$. Moreover, we set the false positive rate 
    bound to $\alpha = 0.05$ and the temperature to 1.
    }
    \label{fig:app-robust}
\end{figure}

In this section, we analyze the robustness of our auditing framework to approximate model access. 
More concretely, we consider a setting in which the auditor has access to a (non-quantized) model, which they use to compute the probabilities $P^\Mcal$ in Algorithm~\ref{alg:test}; however, the provider deploys a quantized version of the model (\texttt{RedHatAI/Llama-3.2-1B-Instruct-FP8}, \texttt{RedHatAI/gemma-3-1b-it-quantized.w8a8}, and \texttt{QuantFactory/Ministral-8B-Instruct-2410-GGUF}) and hence the output token sequences are not sampled according to the exact distribution $P^\Mcal$ used by the provider.
Figure~\ref{fig:app-robust} summarizes the results, which show our auditing framework is indeed robust to such approximate model access.

\clearpage
\newpage

\section{Misreporting Policies}\label{app:policies}

Here, we describe in detail the misreporting policies, first introduced in~\citep{velasco2025llmoverchargingyoutokenization}, that we consider for the experiments in Section~\ref{sec:experiments}. In Algorithm~\ref{alg:heuristic}, given a token sequence $\tb$, we denote by $\Vcal_p(\tb)$ the smallest subset of $\Vcal$ whose cumulative next-token probability is at least $p\in(0,1)$~\citep{Holtzman2020The}.

\setlength{\textfloatsep}{10pt}
\begin{algorithm}[h]
\caption{
It returns a token sequence $\tilde{\tb}$ longer or equal than $\tb$ }
\label{alg:random}
\begin{algorithmic}

\State \textbf{Input} Generated output token sequence $\tb$, number of iterations $m$, LLM vocabulary $\Vcal$

\State \textbf{Initialize} $\tilde{\tb}\gets \tb$

\For{$m$ iterations} 
    \State \texttt{valid\_splits} $\gets \{(i,t_1,t_2)\ \text{such that}\ i \in [\texttt{len}(\tilde{\tb})], t_1, t_2\in \Vcal, \ \text{and}\ \texttt{str}(t_1,t_2)=\texttt{str}(\tilde{t}_i) \}$ 
    \State  $(i,t_1,t_2) \gets \text{Random}(\texttt{valid\_splits})$ 
        \State \textbf{break} \
    \If{|$\texttt{valid\_splits}$|=0}
        \State \textbf{break}
    \EndIf

    \State $\tilde{\tb} \gets \left(\tilde{\tb}_{<i},t_1, t_2, \tilde{\tb}_{>i}\right)$ 
\EndFor
\State \Return $\tilde{\tb}$
\end{algorithmic}
\end{algorithm}

\setlength{\textfloatsep}{10pt}
\begin{algorithm}[h]
\caption{It returns a plausible token sequence $\tilde{\tb}$ 
longer or equal than 
$\tb$
}
\label{alg:heuristic}
\begin{algorithmic}

\State \textbf{Input} True output token sequence $\tb$, number of iterations $m$, top-$p$ sampling parameter $p$, token-to-id function $\texttt{id}(\bullet)$ for the vocabulary $\Vcal$ of the LLM $\Mcal$

\State \textbf{Initialize} $\tb'\gets \tb$


\For{$m$ iterations} 
    \State  $i \gets \argmax_{j\in[\texttt{len}(\tb')]}\texttt{id}(t'_j)$ 
    \If {$|\texttt{str}\left(t'_i\right)| = 1$}
        \State \textbf{break}
    \EndIf
    \State $(t^*_1, t^*_2) \gets \argmax_{v_1, v_2\in\Vcal \,:\, \texttt{str}\left(\left(v_1, v_2\right)\right) = \texttt{str}\left(t'_i\right)} \min\left(\texttt{id}(v_1), \texttt{id}(v_2)\right)$
    \State $\tb' \gets \left(\tb'_{<i},t^*_1, t^*_2, \tb'_{>i}\right)$ 
\EndFor

\If{$t'_i \in \Vcal_p(\tb'_{\leq i-1}) \,\,\forall i \in [ \texttt{len}\left( \tb'\right)],$} 

    \State $\tilde{\tb} \gets \tb'$
    
\Else
    \State $\tilde{\tb} \gets \tb$
    
\EndIf
\State \Return $\tilde{\tb}$
\end{algorithmic}
\end{algorithm}

\clearpage
\newpage

\section{Proofs}\label{app:proofs}
\subsection{Proof of Proposition~\ref{prop:unbiased-estimator}}

The proof of the proposition builds on the literature on debiasing Monte Carlo estimators via randomized truncation, originating as a variance-reduction technique in Monte Carlo simulations~\cite {Kahn1953} and later developed in the context of stochastic differential equations~\citep{rhee2012newapproachunbiasedestimation}.

To prove the proposition, we begin by considering a fixed prompt $q$ and a string $\sbb$ generated by the LLM $\Mcal$ as a response to $q$. We fix a distribution $P^K$ supported over $\mathbb{N}$ and sample $K\sim P^K$. For ease of exposition, we consider $\widehat{\Tb}_k \sim \widehat{P}^\Mcal_{\sbb}(\cdot \given q)$ is defined for each integer $k\geq 1$, and it is given by Eq.~\ref{eq:locally-constrained}. However, note that, to run Algorithm~\ref{alg:estimator}, an auditor only needs to compute the first $K$ samples, \ie, $\widehat{\Tb}_1, \dots, \widehat{\Tb}_K$. 

We will now show that the sequence of estimators used by Algorithm~\ref{alg:estimator} and defined by $R_0=0$ and for $k\geq 1$ by
\begin{equation}
    R_k=
        \frac{\overbrace{\frac{1}{k}\sum_{j=1}^k \frac{P^\Mcal(\widehat{\Tb}_j \given q)}{\widehat{P}^{\Mcal}_{\sbb}(\widehat{\Tb}_j \given q)}\cdot \textnormal{\texttt{len}}(\widehat{\Tb}_{j})}^{\dagger(k)}}{\underbrace{\frac{1}{k}\sum_{j=1}^k \frac{P^\Mcal(\widehat{\Tb}_j \given q)}{\widehat{P}^{\Mcal}_{\sbb}(\widehat{\Tb}_j \given q)}}_{\ddagger(k)}},
\end{equation}
convergences as $k\to\infty$. To this end, we will prove that both $\dagger(k)$ and $\ddagger(k)$ converge separately as $k\to\infty$. Firstly, for the term $\ddagger(k)$, we have have that
\begin{align}
    \lim_{k\to\infty} \ddagger(k) &=\lim_{k\to\infty} \frac{1}{k}\sum_{j=1}^k \frac{P^\Mcal(\widehat{\Tb}_j \given q)}{\widehat{P}^{\Mcal}_{\sbb}(\widehat{\Tb}_j \given q)}\\
    &\overset{(i)}{=} \EE_{\widehat{\Tb}\sim \widehat{P}_{\sbb}^\Mcal(\cdot \given q)}\left[ \frac{P^\Mcal(\widehat{\Tb} \given q)}{\widehat{P}_{\sbb}^\Mcal(\widehat{\Tb} \given q)} \right]\\
    &= \sum_{\hat{\tb}\in\Vcal^* \,\colon\, \texttt{str}(\hat{\tb}) = \sbb} \frac{P^\Mcal(\hat{\tb} \given q)}{\widehat{P}_{\sbb}^\Mcal(\hat{\tb} \given q)}\cdot \widehat{P}_{\sbb}^\Mcal(\hat{\tb} \given q)\\
    &= \sum_{\tb\in\Vcal^* \,\colon\, \texttt{str}(\tb) = \sbb} P^\Mcal(\tb \given q)\\
    &= P^\Mcal(\sbb \given q) ,
\end{align} 
where $(i)$ follows by the Law of Large Numbers, since $\widehat{\Tb}_{k}\sim P_{\sbb}^\Mcal(\cdot \given q)$ for each $k$, and
$P^\Mcal(\sbb \given q)$ denotes the probability that the LLM $\Mcal$ generates the output string $\sbb$.\footnote{Note that, in practice, as LLMs have a finite context window, the sequence of tokens they can generate is finite, and hence, the distribution $P^\Mcal(\cdot \given q)$ can be formally considered to have finite support and finite variance, which is sufficient to apply the Law of Large Numbers.}
%

Secondly, we consider the term $\dagger(k)$ and note that,

\begin{align}
    \lim_{k\to\infty} \dagger(k)&=\lim_{k\to\infty} \frac{1}{k}\sum_{j=1}^k \frac{P^\Mcal(\widehat{\Tb}_j \given q)}{\widehat{P}^{\Mcal}_{\sbb}(\widehat{\Tb}_j \given q)}\cdot \textnormal{\texttt{len}}(\Tb_{j})\\
    &\overset{(ii)}{=}\EE_{\widehat{\Tb}\sim \widehat{P}_{\sbb}^\Mcal(\cdot \given q)} \left[ \frac{P^\Mcal(\widehat{\Tb} \given q)}{\widehat{P}^{\Mcal}_{\sbb}(\widehat{\Tb} \given q)}\cdot \textnormal{\texttt{len}}(\widehat{\Tb}) \right]\\
    &=\sum_{\tb\in\Vcal^* \,\colon\, \texttt{str}(\tb)=\sbb} P^\Mcal(\tb \given q)\cdot \texttt{len}(\tb)\\
\end{align}
where $(ii)$ follows again from the Law of Large Numbers.

As a result, using the above limits for $\dagger(k)$ and $\ddagger(k)$, we can conclude that:
\begin{equation}\label{eq:app-proof-estimator-R}
    \begin{aligned}
    R_\infty\coloneqq\lim_{k\to\infty}R_k &= \frac{\lim_{k\to\infty} \dagger(k)}{\lim_{k\to\infty}\ddagger(k)}\\
    &=\frac{\sum_{\tb\in\Vcal^* \,\colon\, \texttt{str}(\tb)=\sbb}\texttt{len}(\tb)\cdot P^\Mcal(\tb \given q)}{P^\Mcal(\sbb \given q)}\\
    &=\sum_{\tb\in\Vcal^* \,\colon\, \texttt{str}(\tb)=\sbb}\texttt{len}(\tb)\cdot P^\Mcal_{\sbb}(\tb \given q)\\
    &=\EE_{\Tb \sim P^{\Mcal}_{\sbb}(\cdot\given q)}\left[ \textnormal{\texttt{len}}(\Tb)\right]\\
\end{aligned},
\end{equation}
which, in particular, shows that $R_\infty$ is a fixed constant depending exclusively on $q$, $\sbb$ and $\Mcal$.

Finally, the estimator $\sum_{k=1}^K \frac{R_k-R_{k-1}}{P(K\geq k)} $ constructed by Algorithm~\ref{alg:estimator} satisfies

\begin{align}
    \EE_{K\sim P^K,\, \widehat{\Tb}_{k}\sim \widehat{P}^{\Mcal}_{\sbb}} \left[ \sum_{k=1}^K \frac{R_k-R_{k-1}}{P(K\geq k)} \right] &= \EE_{K\sim P^K,\, \widehat{\Tb}_{k}\sim \widehat{P}^{\Mcal}_{\sbb}} \left[ \sum_{k=1}^\infty \mathds{1}\{k\leq K \} \frac{R_k-R_{k-1}}{P(K\geq k)} \right]\\
    &  \overset{(*)}{=}\sum_{k=1}^\infty \EE_{K\sim P^K}\left[\mathds{1}\{K\geq k \} \right]
    \cdot
    \EE_{\widehat{\Tb}_{k}\sim \widehat{P}^{\Mcal}_{\sbb}}\left[\frac{R_k-R_{k-1}}{P(K\geq k)}\right] \\
    & =\sum_{k=1}^\infty P(K\geq k)
    \cdot
    \EE_{ \widehat{\Tb}_{k}\sim \widehat{P}^{\Mcal}_{\sbb}} \left[\frac{R_k-R_{k-1}}{P(K\geq k)} \right]\\
    & =\sum_{k=1}^\infty \EE_{ \widehat{\Tb}_{k}\sim \widehat{P}^{\Mcal}_{\sbb}} \left[R_k-R_{k-1}\right]\\
    & \overset{(**)}{=} \EE_{ \widehat{\Tb}_{k}\sim \widehat{P}^{\Mcal}_{\sbb}}\left[R_\infty\right]\\
    &\overset{(***)}{=}\EE_{\Tb \sim P^{\Mcal}_{\sbb}(\cdot\given q)}\left[ \textnormal{\texttt{len}}(\Tb)\right],
\end{align}
where in $(*)$ we have used that the sequence of random variables $\widehat{\Tb}_k$ is {i.i.d} and is independent of $K$, in $(**)$ we have used that the sum is telescoping, and in $(***)$ we have used Eq.~\ref{eq:app-proof-estimator-R}.

\subsection{Proof of Proposition~\ref{prop:martingale}}
To prove that the process $M$ defined in Eq.~\ref{eq:martignale-definition} is a martingale under $H_0$, we first conclude that:
\begin{align*}
    &\EE_{Q_i \sim P^{Q}, \, \Tb_i \sim P^{\Mcal} } \left[  \EE_{\widetilde{\Tb}_i \sim\pi_0(q_i,\tb_i), \, K_i\sim P^K,\, \widehat{\Tb}_{k, i}\sim \widehat{P}^{\Mcal}_{\texttt{str}(\tb_i)}(\cdot \given q_i)} \big[E_i \, \given \, \Tb_i=\tb_i, Q_i=q_i \big] \right]\\ 
    &\overset{(i)}{=} \EE_{Q_i \sim P^{Q}, \, \Tb_i \sim P^{\Mcal}} \left[  \EE_{K_i\sim P^K,\, \widehat{\Tb}_{k, i}\sim \widehat{P}^{\Mcal}_{\texttt{str}(\tb_i)}} \left[\textnormal{\texttt{len}}(\Tb_i) -\texttt{EstimateLength}(Q_i,\texttt{str}(\Tb_i),P^\Mcal,P^k) \, \given \, \Tb_i=\tb_i, Q_i=q_i \right] \right]\\
    &= \EE_{Q_i \sim P^{Q}, \, \Tb_i \sim P^{\Mcal}} \left[  \texttt{len}(\Tb_i) \right] 
    \\
    &\quad - \EE_{Q_i \sim P^{Q}, \, \Tb_i \sim P^{\Mcal}} \left[  \EE_{K_i\sim P^K,\, \widehat{\Tb}_{k, i}\sim \widehat{P}^{\Mcal}_{\texttt{str}(\tb_i)}} \left[\texttt{EstimateLength}(Q,\texttt{str}(\Tb_i),P^\Mcal,P^k) \given \, \Tb_i=\tb_i, Q_i=q_i \right] \right]\\
    &\overset{(ii)}{=} \EE_{Q_i \sim P^{Q}, \, \Tb \sim P^{\Mcal}} \left[  \texttt{len}(
    \Tb_i) \right] - \EE_{Q_i \sim P^{Q}, \, \Tb \sim P^{\Mcal}} \left[  \EE_{\Tb' \sim P^{\Mcal}_{\texttt{str}(\tb_i)}}\left[ \textnormal{\texttt{len}}(\Tb')\right] \, \given \Tb_i=\tb_i,\, Q_i=q_i \right]\\
    &=\EE_{Q_i \sim P^{Q}, \, \Tb \sim P^{\Mcal}} \left[  \texttt{len}(\Tb_i) \right] - \EE_{Q_i \sim P^{Q}, \, \Tb \sim P^{\Mcal}} \left[  \texttt{len}(\Tb_i) \right]\\
    &= 0,
\end{align*}
where $(i)$ holds because $\widetilde{\Tb}_i = \Tb_i$ since, under $H_0$, the provider is faithful, and $(ii)$ holds because of Proposition~\ref{prop:unbiased-estimator}. 
Then, since the sequence $E_i$ is independent, it holds under $H_0$ that
\begin{align*}
        \EE_{Q_i \sim P^{Q}, \, \Tb_i \sim P^{\Mcal}(\cdot|Q_i), \, \widetilde{\Tb}_i \sim\pi_0(Q_i,\Tb_i)}\left[ M_i \given M_{i-1} \right] &= \EE_{Q_i \sim P^{Q}, \, \Tb_i \sim P^{\Mcal}(\cdot|Q_i), \, \widetilde{\Tb}_i \sim\pi_0(Q_i,\Tb_i)}\left[ M_{i-1} \cdot (1+ \lambda_i\cdot E_i) \given M_{i-1} \right]\\
        &= M_{i-1} \cdot \EE_{Q_i \sim P^{Q}, \, \Tb_i \sim P^{\Mcal}(\cdot|Q_i), \, \widetilde{\Tb}_i \sim\pi_0(Q_i,\Tb_i)}\left[ 1+\lambda_i\cdot E_i  \right]\\
        &=M_{i-1}.
\end{align*}
This concludes the proof.

\subsection{Proof of Theorem~\ref{thm:validity}}

By assumption, we have that $1+\lambda_i\cdot E_i>0$ for all $i \geq 1$.
As an immediate consequence, the process $M$ defined by

\begin{equation*}
    M_i =
        \begin{dcases}
            1 & i=0\\
            M_{i-1} \cdot  (1+\lambda_i \cdot E_i) & i\geq 1
        \end{dcases}
\end{equation*}
is positive. Moreover, under $H_0$,  $M$ is also a martingale by Proposition~\ref{prop:martingale}. As a result, Ville's inequality~\citep{ville1939étude} guarantees that, under $H_0$,

\begin{align*}
    P_{H_0}\left(\phi_\alpha =1 \right) &= P_{H_0}\left( \left\{ \exists i\in\mathbb{N}\, \colon\, M_i > \frac{1}{\alpha} \right\}\right)\\
    &= P_{H_0}\left( \left\{ \sup_{i\in\mathbb{N}} M_i > \frac{1}{\alpha} \right\}\right)\\
    & \leq \frac{\EE_{H_0}[M_1]}{\alpha}\\
    &= \frac{1}{\alpha},
\end{align*}
where the probability $P_{H_0}$ is taken across al variables $Q_i, \Tb_i, \widetilde{\Tb}_i, K_i$ and $\widehat{\Tb}_{i, k}$ appearing in Algorithm~\ref{alg:test}.

\subsection{Proof of Theorem~\ref{thm:consistency}}
\subsubsection{Proof of Part $i)$}
We fix a misreporting policy $\pi$ such that $\Ical(\pi)>0$, and $\lambda_i = \lambda_0 / i$, where $1+\lambda_0\cdot E>0$ under $H_0$. We first note that, since a misreporting policy only increases the length of the reported tokenizations $\widetilde{\Tb}$ compared to $\Tb$, the variable $E$ defined in Eq.~\ref{eq:e-variable-definition} takes higher values under $H_1$. Consequently, under $H_1$, it also holds that $1+\lambda_0\cdot E>0$.

Our first observation is that the probability that the detection time $\tau$ takes a value higher than an integer $n$ satisfies:
\begin{equation}\label{eq:app-consistency-target}
    \sum_{n=1}^\infty  P_{H_1}(\tau\geq n) 
    \leq \sum_{n=1}^\infty P_{H_1}\left(\sum_{i=1}^n\log\left(1+\frac{\lambda_0\cdot E_i}{i}\right) \leq \log 1/\alpha \right),
\end{equation}
because the condition $\tau \geq n$ precisely means that, at time $n$, the process has not yet reached the threshold $1/\alpha$. Building on this observation, the strategy of the proof is to relate the right-hand side of Eq.~\ref{eq:app-consistency-target}, which contains a logarithm, with the misreporting intensity $\Ical(\pi)$.

To this end, consider a bound $B$ on the random variable $\lambda_0\cdot|E|$, and let $i_0\geq 2B$. Then, using the inequality
\begin{equation*}
    |\log(1+x)-x|\leq 2x^2\quad \text{for}\quad |x|\leq 1/2,
\end{equation*}
we obtain for any $i\geq i_0$:
\begin{equation}\label{eq:app-consistency-log-inequality}
    \frac{\lambda_0\cdot|E_i|}{i}\leq \frac{1}{2} \implies \left|\log\left(1+\frac{\lambda_0\cdot E_i}{i}\right)-\frac{\lambda_0\cdot E_i}{i}\right|\leq 2\cdot\frac{\lambda_0^2\cdot E_i^2}{i^2}
\end{equation}
As a result, taking expectations in the above inequality, for any $i\geq i_0$,
\begin{equation}\label{eq:app-consistency-double-inequality}
    \begin{dcases}
        \EE_{H_1}\left[\log\left(1+\frac{\lambda_0\cdot E_i}{i}\right)\right]\geq \frac{\lambda_0\cdot\Ical(\pi)}{i}-2\frac{\EE_{H_1}[\lambda_0^2\cdot E_i^2]}{i^2}\\
        \EE_{H_1}\left[\log\left(1+\frac{\lambda_0\cdot E_i}{i}\right)\right]\leq \frac{\lambda_0\cdot\Ical(\pi)}{i}+2\frac{\EE_{H_1}[\lambda_0^2\cdot E_i^2]}{i^2}
    \end{dcases},
\end{equation}
where the expectations are taken with respect to all random variables appearing in Algorithm~\ref{alg:test}, and we have used that, as a result of Proposition~\ref{prop:unbiased-estimator}:
\begin{equation}
    \EE_{H_1}[E_i] = \Ical(\pi)>0,\quad i\geq1.
\end{equation}
Now, for any $n\geq i_0$, using Eq.~\ref{eq:app-consistency-double-inequality}, we obtain:
\begin{align}
    \sum_{i=1}^n\EE_{H_1}\left[\log\left(1+\frac{\lambda_0\cdot E_i}{i}\right)\right]&= \sum_{i=0}^{i_0-1}\EE_{H_1}\left[\log\left(1+\frac{\lambda_0\cdot E_i}{i}\right)\right]+\sum_{i=i_0}^n \EE_{H_1}\left[\log\left(1+\frac{\lambda_0\cdot E_i}{i}\right)\right]\\
    &\geq \sum_{i=0}^{i_0-1}\EE_{H_1}\left[\log\left(1+\frac{\lambda_0\cdot E_i}{i}\right)\right] +\sum_{i=i_0}^n \frac{\lambda_0\cdot\Ical(\pi)}{i} - 2 \sum_{i=i_0}^n \EE_{H_1}[\lambda_0^2\cdot E_i^2]\cdot\frac{1}{i^2}.
\end{align}
We can now readily related the right-hand-side of Eq.~\ref{eq:app-consistency-target} with $\Ical(\pi)$. Indeed, since the variables $\lambda_0\cdot E_i$ are bounded by $B$, the series $\sum_{i=1}^\infty 1/i^2$ converges, and the harmonic sum behaves as $\sum_{i=1}^n 1/i \geq \log(n+1)$, we can choose a constant $K$ such that, for any $n\geq 1$, we have:
\begin{equation}\label{eq:app-consistency-harmonic}
    \sum_{i=1}^n\EE_{H_1}\left[\log\left(1+\frac{\lambda_0\cdot E_i}{i}\right)\right] \geq \lambda_0\cdot\Ical(\pi)\cdot \log(n+1)-K.
\end{equation}
The above inequality will allow us to derive an explicit bound for the expectation of $\tau$. To this end, combining Eq.~\ref{eq:app-consistency-log-inequality} and Eq.~\ref{eq:app-consistency-double-inequality}, we obtain that for $i\geq i_0$,
\begin{equation}
     \bigg|\underbrace{\log\left(1+\frac{\lambda_0\cdot E_i}{i}\right)-\EE_{H_1}\left[ \log\left(1+\frac{\lambda_0\cdot E_i}{i}\right) \right]}_{\dagger_i}\bigg|\leq \frac{B+\lambda_0\cdot\Ical(\pi)}{i}+2\cdot\frac{B^2+\EE_{H_1}[\lambda_0^2\cdot E_i]^2}{i^2}=\Ocal(1/i).
\end{equation}
As a result, we can choose a sequence $c_i$ that is constant for $1\leq i<i_0$ and that is $c_{i_0} / i$ for $i\geq i_0$ such that $\sum_{i=1}^\infty c_i^2 = c < \infty$ and for any $i\geq 1$,
\begin{equation}\label{eq:app-consistency-c}
     \left|\log\left(1+\frac{\lambda_0\cdot E_i}{i}\right)-\EE_{H_1}\left[ \log\left(1+\frac{\lambda_0\cdot E_i}{i}\right) \right]\right|\leq c_i.
\end{equation}
We can now apply Hoeffding's inequality to the sequence of variables $\dagger_i$ in the left-hand side of Eq.~\ref{eq:app-consistency-c}, which has mean $0$, to obtain for any $n\geq 1$:
\begin{align}\label{eq:app-consistency-hoeffding}
      P_{H_1}\left(\sum_{i=1}^n\log\left(1+\frac{\lambda_0\cdot E_i}{i}\right) \leq \log 1/\alpha \right) &\leq 2\exp\left(-\frac{\left( - \log 1/\alpha + \sum_{i=1}^n \EE_{H_1}[\log(1+\lambda_0\cdot E_i/i)]\right)^2}{\sum_{i=1}^n c_i^2} \right)\\
      &\leq 2\exp\left(-\frac{\left( -\log 1/\alpha + \sum_{i=1}^n \EE_{H_1}[\log(1+\lambda_0\cdot E_i/i)]\right)^2}{c} \right)\notag
\end{align}
We can then obtain:
\begin{align}\label{eq:app-consistency-partial-bound}
    \EE[\tau] &= \sum_{n=1}^\infty  n\cdot P_{H_1}(\tau= n)\notag\\
    &\overset{(*)}{=}\sum_{n=1}^\infty  P_{H_1}(\tau\geq n)\notag \\
    &\leq \sum_{n=1}^\infty P_{H_1}\left(\sum_{i=1}^n\log\left(1+\frac{\lambda_0\cdot E_i}{i}\right) \leq \log 1/\alpha \right)\notag\\
    &\overset{(**)}{\leq} \sum_{n=1}^\infty 2\exp\left(-\frac{\left( -\log 1/\alpha + \sum_{i=1}^n \EE[\log(1+\lambda_i\cdot E_i/i)]\right)^2}{c} \right)\\\notag
\end{align}
where note that $(*)$ is a general standard property that holds for any integer random variable, and in $(**)$ we have used Eq.\ref{eq:app-consistency-hoeffding}.
Lastly, from Eq.~\ref{eq:app-consistency-harmonic}, if $\Ical(\pi)>0$, it follows that there exists an index $j$ such that for $n\geq j$
\begin{equation*}
    -\log 1/\alpha +  \sum_{i=1}^n\EE\left[\log\left(1+\frac{\lambda_0\cdot E_i}{i}\right)\right] \geq -\log 1/\alpha + \lambda_0\cdot\Ical(\pi)\cdot \log(n+1)-K\geq 0,
\end{equation*}
and hence, using Eq.~\ref{eq:app-consistency-partial-bound} we can finally conclude that:

\begin{align}
    \EE_{H_1}[\tau] &\leq \sum_{n=1}^{j-1} 2\exp\left(-\frac{\left( -\log 1/\alpha + \sum_{i=1}^n \EE_{H_1}[\log(1+\lambda_i\cdot E_i/i)]\right)^2}{c} \right)\\
    &+ \sum_{n=j}^{\infty} 2\exp\left(-\frac{\left( -\log 1/\alpha + \sum_{i=1}^n \EE_{H_1}[\log(1+\lambda_i\cdot E_i/i)]\right)^2}{c} \right)\\
    &\leq \sum_{n=1}^{j-1} 2\exp\left(-\frac{\left( -\log 1/\alpha + \sum_{i=1}^n \EE_{H_1}[\log(1+\lambda_i\cdot E_i/i)]\right)^2}{c} \right)\\
    &+ \sum_{n=j}^{\infty} 2\exp\left(-\frac{\left( -\log 1/\alpha + \lambda_0\cdot\Ical(\pi)\cdot \log(n+1)-K\right)^2}{c} \right)\\
    &< \infty
\end{align}
In particular, $\EE_{H_1}[\tau]<\infty $ implies that $P_{H_1}(\tau<\infty)=1$.
This proves part $i)$.

\subsubsection{Proof of Part $ii)$}

We fix a misreporting policy $\pi$ such that $\Ical(\pi)$ and, under $H_1$,

\begin{equation}\label{eq:app-consistency-assumptions}
    \begin{dcases}
         \mathrm{B}_{+}>1+\lambda_0\cdot E>\mathrm{B}_{-}>0\\
          \log(1+\lambda_0\cdot \Ical(\pi)) > \textnormal{Var}(E)\cdot\frac{\lambda_0^2}{2(\mathrm{B}_{-})^2}. 
    \end{dcases}
\end{equation}

We will begin by proving that the above conditions imply that $\EE_{H_1}[\log(1+\lambda_0\cdot E_i)]>0$, which will be of great use later. To this end, we write the second-order Taylor expansion of $\log$ around the point $1+\lambda_0\cdot\Ical(\pi)>1$:
\begin{align}\label{eq:app-consistency-taylor}
    \log (1+\lambda_0\cdot E) =& \log(1+\lambda_0\cdot \Ical(\pi)\\
    &+\log'\left(1+\lambda_0\cdot \Ical(\pi)\right) \cdot \left( 1+\lambda_0\cdot E-1-\lambda_0\cdot \Ical(\pi) \right)\\
    &+\frac{1}{2}\log''(\xi) \cdot \left( 1+\lambda_0\cdot E-1-\lambda_0\cdot \Ical(\pi)  \right)^2,
\end{align}
where $\xi$ is a point in the support of the variable $1+\lambda_0\cdot E$, which satisfies the bounds in Eq.~\ref{eq:app-consistency-assumptions} by assumption. Thus, using that the function $\log''$ is decreasing, and taking expectation in Eq.~\ref{eq:app-consistency-taylor}, we obtain:
\begin{equation}\label{eq:app-consistency-log-bound}
    \EE_{H_1}\left[\log (1+\lambda_0\cdot E) \right] \geq \underbrace{\log(1+\lambda_0\cdot \Ical(\pi)) -\frac{\text{Var}_{H_1}[1+\lambda_0\cdot E]}{2(B_{-})^2}}_{\mu}>0
\end{equation}
We are now in a position to prove Theorem~\ref{thm:consistency}. For that, we first define the following quantities:

\begin{equation*}
    \begin{dcases}
        Y_i \coloneqq \log(1+\lambda_0 \cdot E_i), & i\geq 1\\
        S_n \coloneqq \sum_{i=1}^n Y_i, &n\geq 1.
    \end{dcases}
\end{equation*}
Then, we can write the condition for the detection time $\tau$ as:
\begin{equation*}
    \tau=\inf \{i\, \colon\, M_i >1/\alpha \}=\inf \{i\, \colon\, S_i >\log(1/\alpha) \}.
\end{equation*}
We begin by showing that the detection time $\tau$ is guaranteed to be finite, \ie, $P_{H_1}(\tau<\infty)=1$. To this end, we note that, since the variables $Y_i$ are independent and identically distributed, and by the Strong Law of Large Numbers:
\begin{equation*}
    \frac{1}{n}S_n \to \EE_{H_1}[Y_1]=\EE_{H_1}[\log(1+\lambda_0\cdot E_1)]\overset{(*)}{\geq}\mu>0 \implies S_n \to \infty,
\end{equation*}
where the convergence is almost surely under the distribution $P_{H_1}$, and in $(*)$ we have used Eq.~\ref{eq:app-consistency-log-bound}.
As a consequence,
\begin{equation}
    P_{H_1}\left( \{ \exists n\geq 1 \, \colon \, S_n > \log(1/\alpha)  \} \right)=1 \implies P_{H_1}(\tau<\infty)=1.
\end{equation}
Based on the above, we can bound the expectation for the detection time $\tau$. However, it is important to note that $\tau$, despite being finite ($P_{H_1}(\tau<\infty)=1$), is not necessarily bounded. To proceed, for any $n\geq 1$, we can define the following stopping time:
\begin{equation*}
    \tau \wedge n,
\end{equation*}
where $x\wedge y= \min(x,y)$. The variable $ \tau \wedge n $ is at most $n$, and hence bounded. As a result, we can consider the random variable obtained by evaluating the process $S_n$ at the bounded stopping time $ \tau \wedge n$. This allows the use of Wald's equality to conclude that:
\begin{align}\label{app:consistency-wald}
    \EE_{H_1}\left[ S_{\tau\wedge n}\right] &= \EE_{H_1}\left[ {\tau\wedge n}\right] \cdot \EE_{H_1}\left[ Y_{1}\right]\\
    &\geq \EE_{H_1}\left[ {\tau\wedge n}\right]\cdot \mu,
\end{align}
since the sequence $Y_i$ is independent and identically distributed. To be able to use the above equality, we next show that $S_{\tau\wedge n}$ is bounded. This will allow us to bound $\EE_{H_1}\left[ S_{\tau\wedge n}\right]$, and thus also $\EE_{H_1}\left[ {\tau\wedge n}\right]$. Indeed, we consider two different cases:

\begin{itemize}
    \item Firstly, if $\tau\leq n$, then, by definition of the detection time $\tau$, at time $\tau-1$ the process has not yet reached the threshold $\log(1/\alpha)$, \ie:
\begin{equation*}
    \begin{dcases}
        S_{\tau-1}\leq \log(1/\alpha)\\
        S_{\tau}>\log(1/\alpha),
    \end{dcases}
\end{equation*}
which implies that 
\begin{equation*}
    S_{\tau} = S_{\tau-1}+Y_\tau \leq \log (1/\alpha) + \log B_{+}.
\end{equation*}

\item Secondly, if $\tau>n$, then $S_n\leq\log 1/\alpha$, because the process at time $n$ has not yet reach the threshold $\log(1/\alpha)$.

\end{itemize}
In summary, we have shown that
\begin{equation*}
    \begin{dcases}
        \tau \leq n \implies S_\tau \leq \log(1/\alpha) +\log B_{+}\\
        \tau > n \implies S_\tau \leq \log(1/\alpha)
    \end{dcases}
    \implies S_{\tau\wedge n}\leq \log(1/\alpha)+\log B_{+}
\end{equation*}
Using the above bound in Eq.~\ref{app:consistency-wald}, we obtain:
\begin{equation*}
    \EE_{H_1}\left[ {\tau\wedge n} \right]\leq \frac{\EE_{H_1}\left[ S_{\tau\wedge n}\right]}{\mu}\leq \frac{\log(1/\alpha)+\log B_{+}}{\mu}.
\end{equation*}
Lastly, we note that $\tau\wedge n$ is a finite stopping time, that $\tau\wedge 1, \tau\wedge 2,\dots$ is a monotonically increasing sequence of random variables, and that $\tau\wedge n \to \tau$ as $n\to\infty$. This allows us to conclude that the expectation of the detection time $\tau$ satisfies:
\begin{align}\label{app:consistency-bound-assumption}
        \EE_{H_1}\left[ {\tau} \right] = \lim_{n\to\infty}\EE_{H_1}\left[ {\tau\wedge n} \right]&\leq \frac{\log(1/\alpha)+\log B_{+}}{\mu}\\
        &= \frac{\log(1/\alpha)+ B_{+}}{ \EE_{H_1}[\log(1+\lambda\cdot E)]}\\
        &\overset{(*)}{\leq} \frac{\log(1/\alpha)+ B_{+}}{\log(1+\lambda_0\cdot \Ical(\pi)) -\frac{\textnormal{Var}_{H_1}[1+\lambda_0\cdot E]}{2(B_{-})^2}}\\
        &\leq \frac{\log(1/\alpha)+ B_{+}}{\log(1+\lambda_0\cdot \Ical(\pi)) -\lambda_0^2\cdot\frac{\textnormal{Var}_{H_1}[ E]}{2(B_{-})^2}}.
\end{align}
where in $(*)$ we have used Eq.~\ref{eq:app-consistency-log-bound}. This proves the result.

\end{document}